\documentclass[preprint, review, 9pt]{elsarticle}

\usepackage{geometry, setspace, float}
\usepackage{amssymb}
\usepackage{amsmath}
    \allowdisplaybreaks

\usepackage{lineno}

\biboptions{sort&compress}

\usepackage{newtxtext, newtxmath}
\usepackage[dvipsnames]{xcolor}
\usepackage{graphicx}
\usepackage{dcolumn} % Align table columns on decimal point
\usepackage{xpatch}
    \makeatletter \xpretocmd \start@align{\linenomathWithnumbers}{}{\fail}
\usepackage{mathtools, url}
\usepackage[toc]{appendix}
\usepackage[ruled, linesnumbered]{algorithm2e}
\usepackage[unicode=true, bookmarks=true, bookmarksnumbered=false, bookmarksopen=false, breaklinks=false, pdfborder={0 0 1}, backref=false, colorlinks=false, hidelinks]{hyperref}
    \hypersetup{linkcolor=black, urlcolor=black, citecolor=black, pdfstartview={FitH}}
\usepackage[ruled, linesnumbered]{algorithm2e}

\usepackage[math]{cellspace}
\usepackage{newunicodechar}
    \newunicodechar{，}{\makebox[1em][l]{,}}
    \newunicodechar{。}{\makebox[1em][l]{.}}
    \newunicodechar{：}{\makebox[1em][l]{:}}
    \newunicodechar{；}{\makebox[1em][l]{;}}
    \newunicodechar{（}{(}
    \newunicodechar{）}{)}
    \newunicodechar{“}{``}
    \newunicodechar{”}{''}

\DeclareSymbolFont{CMlargesymbols}{OMX}{cmex}{m}{n}
\let\sumop\relax\let\prodop\relax
\DeclareMathSymbol{\sumop}{\mathop}{CMlargesymbols}{"50}
\DeclareMathSymbol{\prodop}{\mathop}{CMlargesymbols}{"51}

\SetSymbolFont{operators}{normal}{OT1}{ntxtlf}{m}{n}
\SetSymbolFont{operators}{bold}{OT1}{ntxtlf}{b}{n}

\renewcommand{\vec}[1]{\boldsymbol{#1}}
\newcommand{\tensor}[1]{\boldsymbol{#1}}

\newcommand{\p}{\partial}
\newcommand{\dif}{\mathop{}\!\mathrm{d}}
\newcommand{\bn}{\vec{\nabla}}
\renewcommand{\widebar}[1]{\mskip.5\thinmuskip\overline{\mskip-.5\thinmuskip {#1} \mskip-.5\thinmuskip}\mskip.5\thinmuskip}
\newcommand{\ee}{\mathrm{e}}
\newcommand{\ii}{\mathrm{i}}

\DeclareMathOperator*{\diag}{diag}
\DeclareMathOperator*{\poly}{poly}
\DeclareMathOperator*{\Tr}{Tr}
\renewcommand{\ge}{\geqslant}
\renewcommand{\le}{\leqslant}
\newcommand{\ket}[1]{| #1 \rangle}
\newcommand{\bra}[1]{\langle #1 |}
\newcommand{\T}{^{\mathrm{T}}}

\newcommand{\dt}{\delta_t}
\newcommand{\norm}[1]{\| #1 \|}

\allowdisplaybreaks

\begin{document}

\begin{frontmatter}

\title{Toward end-to-end quantum simulation of rapidly distorted turbulence}
% \title{Quantum computing of turbulence using rapid distortion theory}

\author[fir]{Zhaoyuan Meng}

\author[sec]{Leyu Chen}

\author[thi,fou]{Jin-Peng Liu\corref{cor1}}
\ead{liujinpeng@mail.tsinghua.edu.cn}

\author[fir,fiv]{Guowei He\corref{cor1}}
\ead{hgw@lnm.imech.ac.cn}

\cortext[cor1]{Corresponding authors.}

\address[fir]{State Key Laboratory of Nonlinear Mechanics, Institute of Mechanics, Chinese Academy of Sciences, Beijing 100190, PR China}
\address[sec]{LMIB and School of Mathematical Sciences, Beihang University, Beijing 100191, China}
\address[thi]{Yau Mathematical Sciences Center, Tsinghua University, Beijing 100084, China}
\address[fou]{Yanqi Lake Beijing Institute of Mathematical Sciences and Applications, Beijing 100407, China}
\address[fiv]{School of Engineering Sciences, University of Chinese Academy of Sciences, Beijing, 100049, China}

\begin{abstract}
We propose an end-to-end quantum algorithm to simulate rapidly distorted turbulence via linear combination of Hamiltonian (LCHS).
The algorithm comprises three primary stages: the efficient preparation of an initial turbulent state with a prescribed energy spectrum, its subsequent time evolution via LCHS, and the direct measurement of key turbulence statistics.
Our analysis indicates that the algorithm can offer a practical quantum speedup over the classical simulation methods for a sufficiently large computational grid.
We evaluate the quantum resource requirements for simulating a minimal instance of non-trivial turbulence with classical validation.
The numerical results show excellent agreement with ground-truth solutions, capturing both the qualitative evolution of turbulent fields and the quantitative behavior of statistics, including the Reynolds stresses and the fluctuating velocity spectrum.
Despite its linearity, rapidly distorted turbulence captures essential turbulence mechanisms and may inform the development of quantum algorithms for the Navier-Stokes equations.
Our work establishes a foundation for addressing more complex turbulent phenomena on future fault-tolerant quantum computers.
\end{abstract}


\end{frontmatter}

% \linenumbers

\section{Introduction}
Turbulence, a long-standing challenge in classical physics, is a ubiquitous phenomenon of fundamental importance in science and engineering~\cite{Sreenivasan1999_Fluid}.
The central difficulty of its study is rooted in its inherent complexity, which includes chaotic dynamics, the coexistence of ordered and disordered motions, and multi-scale structures, precluding an accurate description and prediction.
Although direct numerical simulation provides high-fidelity solutions~\cite{Rogallo1984_Numerical, Moin1998_Direct, Ishihara2009_Study}, its prohibitive computational cost renders the accurate simulation of high-Reynolds-number turbulence a formidable challenge~\cite{Kim2024_The}.
This computational bottleneck therefore necessitates the development of novel paradigms to overcome this long-standing problem.

The novel paradigm of quantum computing~\cite{Feynman1982_Simulating} holds the potential to address the aforementioned challenges.
By leveraging quantum superposition and entanglement, it enables massively parallel computations, offering a potential exponential advantage over classical computers for specific problems.
The disruptive capability of quantum algorithms has been demonstrated in fields such as cryptography~\cite{Shor1997_Polynomial, Hallgren2007_Polynomial}, solving linear systems~\cite{Harrow2009_Quantum, Gilyen2019_Quantum, Subaşı2019_Quantum, Lin2020_Optimal, Costa2022_Optimal}, and quantum simulation~\cite{Feynman1982_Simulating, Lloyd1996_Universal, Buluta2009_Quantum, Cirac2012_Goals, Georgescu2014_Quantum}.
However, a fundamental theoretical obstacle emerges when applying this approach to fluid dynamics, particularly turbulence simulation.
The Navier-Stokes (NS) equations governing turbulence are strongly nonlinear, whereas the evolution of a closed quantum system is governed by the linear Schrödinger equation, which dictates a unitary and therefore linear transformation of the quantum state.
This fundamental mismatch between a linear computational framework and nonlinear physical dynamics constitutes the central challenge for applying quantum computing to fluid mechanics~\cite{Succi2023_Quantum, Liu2023_Quantum, Meng2023_Quantum, Meng2024_Simulating, Tennie2025_Quantum, Bharadwaj2025_Towards, Meng2025_Challenges, Wang2025_Quantum, Wang2025_Simulating, Meng2025_Advances}.

To resolve this fundamental conflict, several exploratory avenues have been proposed.
These efforts have predominantly centered on mathematical linearizations of the nonlinear NS dynamics, including Carleman embedding~\cite{Liu2021_Efficient, Sanavio2024_Three, Gonzalez-Conde2025_Quantum}, the Fokker-Planck equation~\cite{Tennie2024_Solving}, the Koopman-von Neumann representation~\cite{Joseph2020_Koopman}, the Koopman operator~\cite{Giannakis2022_Embedding, Zhang2025_Data}, the Liouville equation~\cite{Jin2023_Time, Succi2024_Ensemble}, the homotopy analysis method~\cite{Xue2025_Quantum}, and lattice gas cellular automata~\cite{Yepez2001_Quantum}.
Although these approaches offer a theoretical framework for linearizing nonlinear dynamics in a higher-dimensional space, they are general mathematical tools that do not necessarily incorporate physical insights from fluid mechanics.
For turbulent flows of practical interest, these methods require lifting to overwhelming high dimensions, thereby limiting their near-term applicability.
% In practical regimes, however, strong nonlinearities necessitate lifting to a space of intractably high dimension, rendering such methods infeasible for real-world applications.
The present work therefore pursues an alternative, physically-grounded approach by identifying turbulence models whose dominant dynamics are inherently linear under well-defined assumptions.

\begin{figure}[t!]
    \centering
    \includegraphics{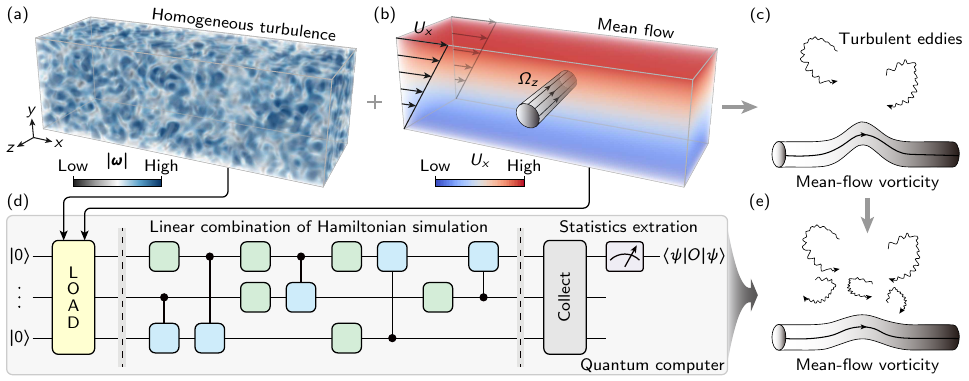}
    \caption{Schematic of the quantum simulation of turbulence using RDT.
    (a) An initial field of homogeneous turbulence is subjected to (b) an instantaneous, strong shear flow characterized by a mean velocity $U_x$ and mean vorticity $\varOmega_z$.
    This scenario corresponds to the rapid distortion limit, where the strong mean shear, $\varOmega_z\gg |\vec{\omega}|$, acts on a timescale much shorter than the intrinsic nonlinear timescale of the turbulence.
    (c) Consequently, an initially straight mean-vortex tube is distorted by the turbulent velocity fluctuations, leading to the formation of a kink.
    (d) An end-to-end quantum algorithm for simulating the non-unitary RDT dynamics via a linear combination of Hamiltonian simulations.
    (e) The kink constitutes a new, small-scale, locally rotating structure, corresponding to the generation of new fluctuation vorticity.
    Through this mechanism, kinetic energy is extracted from the mean flow and transferred to the turbulent fluctuations, as a primary mechanism for turbulence production.}
    \label{fig:schematic_QRDT}
\end{figure}

Accordingly, this work leverages rapid distortion theory (RDT) to establish a connection between linear quantum algorithms and the physics of nonlinear turbulence.
RDT describes the evolution of homogeneous turbulence subjected to a rapid, uniform mean strain, as depicted in Figs.~\ref{fig:schematic_QRDT}(a) and (b).
Its central assumption is that the distortion timescale is much shorter than the intrinsic eddy-turnover time of the turbulence.
Consequently, nonlinear fluctuation-fluctuation interactions and viscous dissipation are negligible, thereby linearizing the governing equations for the turbulent fluctuations.
Despite this linearization, RDT is far from an oversimplification, as its predictions show remarkable agreement with simulations and experiments for homogeneously sheared turbulence, especially during its initial evolution~\cite{Gence1979_On, Britter1979_The, Mann1994_The}.
The theory successfully explains key phenomena such as the generation of anisotropy, the growth of Reynolds shear stress, and the tilting of coherent structures (see Figs.~\ref{fig:schematic_QRDT}(c) and (e)).
Thus, RDT remains a highly effective and widely-used framework in both academia and industry, with applications from turbomachinery to aerospace design~\cite{Terry2000_Suppression, Jacobs2001_Simulations}.
It therefore provides a fundamental basis for understanding how linear, mean-strain-driven mechanisms dominate the dynamics of turbulent structures.
Consequently, RDT captures essential turbulence mechanisms and may inform the development of quantum algorithms for the NS equation.

The evolution of RDT is a high-dimensional problem, as a large number of Fourier modes are required to resolve the fine structure of turbulence, which aligns with the strengths of quantum computing.
A direct quantum implementation of RDT is impeded by its non-unitary dynamics, which conflicts with the unitary evolution required by a quantum computer.
This non-unitarity can be efficiently addressed by two mathematically related techniques, Schrödingerisation~\cite{Jin2023_Quantum, Jin2024_Quantum, Jin2024_Quantum_FP, Jin2024_Quantum_Maxwell, Lu2024_Quantum, Jin2025_On} and the linear combination of Hamiltonian simulation (LCHS)~\cite{An2023_Linear, An2023_Quantum, Yang2025_Circuit, Lu2025_Infinite, Huang2025_Fourier, Wang2025_QuantumCBMD, Low2025_Optimal}.
The former constructs an equivalent unitary evolution by embedding the non-unitary operator within an augmented Hilbert space.
The latter decomposes the target operator into a weighted sum of efficiently implementable unitary operations.
In this work, we employ the LCHS method to perform the non-unitary time evolution.
Moreover, as the performance of a quantum computation depends as critically on state preparation and measurement as on the dynamical evolution, we develop efficient quantum algorithms for these initial and final stages, which form an end-to-end framework for turbulence simulation designed to achieve a holistic quantum speedup.

Crucially, we quantify the minimal resource overhead for simulating non-trivial turbulence, specifically rapidly distorted turbulence, on a quantum computer.
We estimate the qubit count and circuit depth dictated by Trotterization.
Consequently, we establish a fundamental benchmark for the quantum computational cost of fluid dynamics, advancing the field from theoretical algorithms to concrete resource estimations for practical problems.

The paper is organized as follows.
Section~\ref{sec:LCHS_for_RDT} establishes the application of the linear combination of Hamiltonian simulation to rapid distortion theory.
Section~\ref{sec:quantum_algorithm} then details the end-to-end quantum algorithm for turbulence simulation, encompassing initial preparation, evolution, and statistical measurement.
In Sec.~\ref{sec:results}, we present a numerical validation of the proposed framework using a case of rapidly distorted turbulence.
Finally, Sec.~\ref{sec:conclusions} provides the conclusions.

\section{Linear combination of Hamiltonian simulation for rapidly distorted turbulence}\label{sec:LCHS_for_RDT}
\subsection{Rapid distortion theory for homogeneous shear flows}
Consider an incompressible statistically homogeneous turbulence with a linear mean velocity field
\begin{equation}\label{eq:mean_vel}
    U_\alpha = A_{\alpha\beta} x_\beta,
\end{equation}
and a corresponding mean vorticity $\varOmega_\alpha=\varepsilon_{\alpha\beta\gamma} A_{\gamma\beta}$.
The fluctuating velocity is expanded in the Fourier series
\begin{equation}\label{eq:vel_Fourier_expand}
    u_i(\vec{x}, t) = \sum_{\vec{\kappa}(t)} \hat{u}_i(\vec{\kappa}(t), t) \ee^{\ii \vec{\kappa}(t)\cdot\vec{x}}.
\end{equation}
The wavenumber $\vec{\kappa}(t)$ is distorted by the mean flow and consequently evolves in time, with the initial condition $\vec{\kappa}(0)=\vec{k}$.
Substituting Eq.~\eqref{eq:vel_Fourier_expand} and the mean velocity from Eq.~\eqref{eq:mean_vel} into the linearized, inviscid momentum equation
\begin{equation}
    \p_t\vec{u} + \vec{U}\cdot\bn \vec{u} + \vec{u}\cdot\bn\vec{U} = -\bn p
\end{equation}
with constant density $\rho=1$ yields
\begin{equation}\label{eq:RDT_eq_0}
    \dif_t\hat{\vec{u}} + \ii (\vec{x}\cdot \dif_t \vec{\kappa}) \hat{\vec{u}} + \ii (\vec{\kappa}\cdot\tensor{A} \cdot\vec{x}) \hat{\vec{u}} + \tensor{A}\cdot\hat{\vec{u}} = -\ii \hat{p} \vec{\kappa}.
\end{equation}
To eliminate the explicit spatial dependence in Eq.~\eqref{eq:RDT_eq_0}, the wavevector must evolve according to
\begin{equation}\label{eq:kappa_eq}
    \dif_t\vec{\kappa} = -\vec{\kappa}\cdot\tensor{A}.
\end{equation}
This condition simplifies the momentum equation to
\begin{equation}\label{eq:RDT_eq_1}
    \dif_t\hat{\vec{u}} + \tensor{A}\cdot\hat{\vec{u}} = -\ii \hat{p} \vec{\kappa}.
\end{equation}
Projecting Eq.~\eqref{eq:RDT_eq_1} onto $\hat{\vec{u}}$, using the divergence-free condition $\vec{\kappa}\cdot\hat{\vec{u}}=0$ and Eq.~\eqref{eq:kappa_eq}, we obtain the pressure
\begin{equation}\label{eq:pressure}
    \hat{p} = \frac{2\ii \vec{\kappa}\cdot \tensor{A} \cdot \hat{\vec{u}}}{|\vec{\kappa}|^2}.
\end{equation}
Finally, substituting Eq.~\eqref{eq:pressure} into Eq.~\eqref{eq:RDT_eq_1} yields the RDT equation~\cite{Durbin2011_Statistical}
\begin{equation}\label{eq:RDT_eq}
    \dif_t \hat{u}_\alpha = \hat{u}_\beta A_{\gamma\beta}\bigg(\frac{2\kappa_\alpha \kappa_\gamma}{|\vec{\kappa}|^2} - \delta_{\alpha\gamma} \bigg).
\end{equation}

\subsection{Linear combination of Hamiltonian simulation for the RDT equation}
We employ the LCHS technique to cast the RDT equation \eqref{eq:RDT_eq} into a unitary evolution suitable for implementation on a quantum computer.
The formal solution to Eq.~\eqref{eq:kappa_eq} with the prescribed initial condition $\vec{\kappa}(0)=\vec{k}$ is
\begin{equation}\label{eq:kappa(t)}
    \vec{\kappa}(t) = \mathcal{T}\ee^{-\int_0^t \vec{A}\T(s) \dif s}\vec{k},
\end{equation}
where $\mathcal{T}$ is the time-ordering operator, which rearranges a string of operators according to their time arguments, from the latest on the left to the earliest on the right. When $\vec{A}$ is a constant matrix, the formal solution for $\vec{\kappa}(t)$ simplifies to $\vec{\kappa}(t)=\ee^{-\vec{A}\T t}\vec{k}$.
Note that regardless of whether $\vec{A}$ is diagonalizable, the Jordan normal form can be used to compute $\ee^{-\vec{A}\T t}$. When it is not diagonalizable, secular terms such as $t\ee^{\lambda t}$ or $t^2\ee^{\lambda t}$ may appear in the expression for $\ee^{-\vec{A}\T t}$.

Let
\begin{equation}
    M_{\alpha\beta}(\vec{\kappa}) = A_{\gamma\beta} \left(\frac{2\kappa_\alpha\kappa_\gamma}{|\vec{\kappa}|^2} - \delta_{\alpha\gamma} \right).
\end{equation}
Let $N$ be the total number of grid points, and denote $\vec{\mathcal{U}}=[\hat{\vec{u}}(\vec{\kappa}_1), \hat{\vec{u}}(\vec{\kappa}_2), \cdots, \hat{\vec{u}}(\vec{\kappa}_N)]\T$. 
An equivalent form of the RDT equation \eqref{eq:RDT_eq} is
\begin{equation}\label{eq:RDT_eq0}
    \p_t\vec{\mathcal{U}}(t) =  \vec{B}(t)\vec{\mathcal{U}}(t),
\end{equation}
where $\vec{\mathcal{U}} \in \mathbb{C}^{3N}$ and
\begin{equation}
    \vec{B}(t) = \sum_{i=1}^N \ket{\vec{\kappa}_i}\bra{\vec{\kappa}_i} \otimes \vec{M}(\vec{\kappa}_i)
\end{equation}
is a $3N\times 3N$ block-diagonal real matrix. 

Subsequently, $\vec{B}(t)$ is decomposed into its Hermitian and anti-Hermitian parts as
\begin{equation}
    \vec{B}(t) = \vec{L}(t) + \ii\vec{H}(t),
\end{equation}
where $\vec{L}(t) = (\vec{B}(t)+\vec{B}^\dagger(t))/2$ and $\vec{H}(t) = (\vec{B}(t)-\vec{B}^\dagger(t))/(2\ii)$ are both Hermitian matrices.
Let
\begin{equation}
    c = \max_{0\le t\le T} \lambda_{\max}(\vec{L}(t))
\end{equation}
be the maximum eigenvalue of
\begin{equation}\label{eq:def_L}
    \vec{L} = \sum_{\vec{\kappa}} \ket{\vec{\kappa}}\bra{\vec{\kappa}} \otimes \vec{L}_d(\vec{\kappa})
\end{equation}
over the time interval of interest $t\in[0,T]$, where
\begin{equation}\label{eq:L_d}
    \vec{L}_d(\vec{\kappa})
    = \frac{\vec{\kappa}\vec{\kappa}\T\vec{A} + \vec{A}\T\vec{\kappa}\vec{\kappa}\T}{|\vec{\kappa}|^2} - \frac{\vec{A} + \vec{A}\T}{2}.
\end{equation}
Since $\vec{L}$ is a block-diagonal matrix, its maximum eigenvalue is the maximum of the eigenvalues of its diagonal blocks, i.e.,
\begin{equation}
    c = \max_{0\le t\le T}\{\lambda_{\max}(\vec{L}_d(\vec{\kappa}_1)), \lambda_{\max}(\vec{L}_d(\vec{\kappa}_2)), \cdots, \lambda_{\max}(\vec{L}_d(\vec{\kappa}_N)) \}.
\end{equation}
The largest eigenvalue of $\tensor{L}_d(\vec{\kappa})$ has an analytical solution
\begin{equation}
    \lambda_{\max}(\vec{L}_d(\vec{\kappa}))
    = \frac{\Tr(\vec{L}_d(\vec{\kappa}))}{3} + 2\sqrt{\frac{\Tr^2(\vec{L}_d(\vec{\kappa}))}{9} - \frac{C_2(\vec{L}_d(\vec{\kappa}))}{3}}\cos\bigg(\frac{1}{3}\arccos\bigg(-\frac{p/2}{\sqrt{-q^3/27}} \bigg) \bigg),
\end{equation}
where $C_2(\vec{L}_d(\vec{\kappa}))$ is the sum of the principal minors, and the coefficients are
\begin{equation}
    p = C_2(\vec{L}_d(\vec{\kappa})) - \frac{\Tr^2(\vec{L}_d(\vec{\kappa}))}{3}, \quad
    q = -\det(\vec{L}_d(\vec{\kappa})) + \frac{C_2(\vec{L}_d(\vec{\kappa}))\Tr(\vec{L}_d(\vec{\kappa}))}{3}
    - \frac{2\Tr^3(\vec{L}_d(\vec{\kappa}))}{27}.
\end{equation}
According to Weyl's inequality and the Cauchy-Schwarz inequality, we estimate
\begin{equation}
    \lambda_{\max}(\vec{L}_d(\vec{\kappa})) \le 3\sigma_{\max}(\vec{A}),
\end{equation}
where $\sigma_{\max}(\vec{A})$ is the largest singular value or the spectral norm (operator 2-norm) of $\vec{A}$. 
Note that it is independent of $\vec{\kappa}$, and therefore $c\le 3\sigma_{\max}(\vec{A})$.
Consequently, setting $c=3\sigma_{\max}(\vec{A})$ provides a rigorous and sufficiently large upper bound for all wavenumbers $\vec{\kappa}$ and time $t\in [0, T]$.

By the transformation
\begin{equation}
    \tilde{\vec{\mathcal{U}}}=\ee^{-ct}\vec{\mathcal{U}}, \quad
    \tilde{\vec{B}}(t)=\vec{B}(t) - c\tensor{I},
\end{equation}
Eq.~\eqref{eq:RDT_eq0} is transformed into
\begin{equation}\label{eq:RDT_eq1}
    \p_t \tilde{\vec{\mathcal{U}}}(t)
    = \tilde{\vec{B}}(t)\tilde{\vec{\mathcal{U}}}(t),
\end{equation}
where $\tilde{\vec{B}}(t)$ is always semi-negative definite, satisfying the conditions of the LCHS theorem~\cite{An2023_Linear, An2023_Quantum}.
Thus, the time-integration operator for Eq.~\eqref{eq:RDT_eq1} is given by~\cite{An2023_Quantum}
\begin{equation}\label{eq:int_opt_0}
    \mathcal{T}\ee^{\int_0^t \tilde{\vec{B}}(s)\dif s}
    = \int_{\mathbb{R}} \frac{\ee^{2^\beta - (1+\ii r)^{\beta}}}{2\pi (1-\ii r)} \mathcal{T} \ee^{\ii\int_0^t [\vec{H}(s) + r\tilde{\vec{L}}(s)]\dif s} \dif r,
\end{equation}
where $\tilde{\vec{L}}(t)=\vec{L}(t)-c\tensor{I}$ and the constant $\beta\in(0,1)$ governs the decay rate of the kernel function.
Note that each term
\begin{equation}
    \vec{U}_r(t) := \mathcal{T}\ee^{\ii\int_0^t [\vec{H}(s) + r\tilde{\vec{L}}(s)]\dif s}
\end{equation}
in Eq.~\eqref{eq:int_opt_0} is a unitary operator. 
The RDT time evolution is therefore precisely a linear combination of infinitely many unitary operators.

However, in practical implementation, the integral in Eq.~\eqref{eq:int_opt_0} needs to be truncated and discretized into a linear combination of a finite number of unitary operators. 
We truncate the integral over $r$ to the interval $[-R,R]$, which gives
\begin{equation}\label{eq:int_opt_1}
    \tilde{\vec{\mathcal{U}}}(t)
    = \int_{-R}^R \frac{\ee^{2^\beta - (1+\ii r)^{\beta}}}{2\pi (1-\ii r)} \vec{U}_r(t) \tilde{\vec{\mathcal{U}}}_0 \dif r + \vec{E}_{\mathrm{trunc}}^{(1)}.
\end{equation}
We now proceed to estimate the error term
\begin{equation}
    \vec{E}_{\mathrm{trunc}}^{(1)} = \bigg(\int_{-\infty}^{-R} + \int_{R}^\infty \bigg) \frac{\ee^{2^\beta - (1+\ii r)^{\beta}}}{2\pi (1-\ii r)} \vec{U}_r(t) \tilde{\vec{\mathcal{U}}}_0 \dif r.
\end{equation}
By the triangle inequality, we have
\begin{equation}
    \norm{\vec{E}_{\mathrm{trunc}}^{(1)}}_2
    \le \int_{|r|\ge R} \bigg\| \frac{\ee^{2^\beta - (1+\ii r)^{\beta}}}{2\pi (1-\ii r)} \vec{U}_r(t) \tilde{\vec{\mathcal{U}}}_0 \bigg\|_2 \dif r.
\end{equation}
Since $\tensor{U}_r(t)$ is a unitary operator, i.e. $\norm{\vec{U}_r(t) \tilde{\vec{\mathcal{U}}}_0}_2=\norm{\tilde{\vec{\mathcal{U}}}_0}_2$, we therefore obtain
\begin{equation}\label{eq:E_trun_1}
    \norm{\vec{E}_{\mathrm{trunc}}^{(1)}}_2
    \le \norm{\tilde{\vec{\mathcal{U}}}_0}_2 \bigg|\int_{|r|\ge R} \frac{\ee^{2^\beta - (1+\ii r)^{\beta}}}{2\pi (1-\ii r)} \dif r \bigg|
    \le \frac{\ee^{2^{\beta} - R^{\beta}\cos(\frac{\pi\beta}{2})}}{\pi \beta R^{\beta} \cos(\frac{\pi\beta}{2})}  \norm{\tilde{\vec{\mathcal{U}}}_0}_2 
    \lesssim O(\ee^{-R^\beta}).
\end{equation}
Substituting Eq.~\eqref{eq:E_trun_1} into Eq.~\eqref{eq:int_opt_1} yields
\begin{equation}
    \tilde{\vec{\mathcal{U}}}(t)
    = \int_{-R}^R \frac{\ee^{2^\beta - (1+\ii r)^{\beta}}}{2\pi (1-\ii r)} \vec{U}_r(t) \tilde{\vec{\mathcal{U}}}_0 \dif r + O(\ee^{-R^\beta}).
\end{equation}

The integral is then discretized by partitioning the interval $[-R,R]$ into $I_R = 2R/h$ subintervals $[\ell h, (\ell+1)h]$, for which $R/h$ is an integer and $\ell=-R/h, \cdots, R/h-1$.
A $Q$-point Gaussian quadrature is then applied within each subinterval~\cite{An2023_Quantum}.
This composite quadrature rule yields
\begin{equation}\label{eq:Ut_error2}
    \tilde{\vec{\mathcal{U}}}(t) = \sum_{\ell=-R/h}^{R/h-1} \sum_{q=0}^{Q-1} c_{q,\ell} \vec{U}_{r_{q,\ell}}(t)\tilde{\vec{\mathcal{U}}}_0 + \vec{E}_{\mathrm{trunc}}^{(2)} + O(\ee^{-R^\beta}).
\end{equation}
Here, the Gaussian nodes are located at $r_{q,\ell} = h s_q / 2 + (2\ell + 1)h / 2$, where $s_q$ are the roots of the $Q$th-order Legendre polynomial $P_Q(s)$.
The associated weights are $c_{q,\ell} = w_q \ee^{2^\beta - (1+\ii r_{q,\ell})^{\beta}}/(2\pi (1 - \ii r_{q,\ell}))$, with the Gaussian weights $w_q = h/[(1 - s_q^2) {P_Q'}^2(s_q))$ being independent of the subinterval index $\ell$.
% where the coefficients are $c_j=2Rw_j \ee^{2^\beta - (1+\ii r_j)^{\beta}}/(2\pi M (1 - \ii r_j))$, with $r_j=-R+2jR/M$ and $w_j=1-(\delta_{j,0}+\delta_{j,M})/2$ being the weights for the trapezoidal rule.
For notational simplicity, we collapse the double summation $\sum_{\ell=-R/h}^{R/h-1} \sum_{q=0}^{Q-1}$ into a single sum $\sum_{j=0}^{M-1}$, where $M=I_R Q$.
The truncated error in Eq.~\eqref{eq:Ut_error2} is then estimated by~\cite{An2023_Quantum}
\begin{equation}
    \vec{E}_{\mathrm{trunc}}^{(2)}
    = \bigg\|\int_{-R}^R \frac{\ee^{2^\beta - (1+\ii r)^{\beta}}}{2\pi (1-\ii r)} \vec{U}_r(t) \tilde{\vec{\mathcal{U}}}_0 \dif r - \sum_{j=0}^{M-1} c_j \vec{U}_{r_j}(t)\tilde{\vec{\mathcal{U}}}_0 \bigg\|_2
    \le \frac{4\ee^{2\beta}}{3\pi}R \bigg(\frac{\ee h t\norm{\tensor{L}(t)}_\infty}{2} \bigg)^{2Q} \norm{\tilde{\vec{\mathcal{U}}}_0}_2.
\end{equation}
For the short-time evolution in rapidly distorted turbulence, under the assumption that $\norm{\tensor{L}(t)}_\infty = O(1)$, the discretized propagator $\tilde{\vec{\mathcal{U}}}(t)$ becomes
\begin{equation}\label{eq:Ut_0}
    \tilde{\vec{\mathcal{U}}}(t) = \sum_{j=0}^{M-1} c_{j} \vec{U}_{r_{j}}(t)\tilde{\vec{\mathcal{U}}}_0 + O(\ee^{-R^\beta}, R(\ee ht/2)^{2Q}).
\end{equation}

To evaluate the unitary operators in Eq.~\eqref{eq:Ut_0}, the total evolution time $t$ is discretized into $I_t$ uniform steps of duration $\delta_t=t/I_t$.
A $Q_t$-point Gaussian quadrature is then applied to the time integral within each subinterval $[n_t\delta_t, (n_t+1)\delta_t]$, for $n_t=0,1,\cdots,I_t-1$.
This temporal discretization yields the approximation
\begin{equation}\label{eq:time_evo_opt_0}
    \vec{U}_r(t) \tilde{\vec{\mathcal{U}}}_0
    = \prod_{n_t=0}^{I_t-1} \exp\bigg( \ii \sum_{q_t=0}^{Q_t-1} w_{q_t} \tilde{\tensor{B}}((2n_t+1)\delta_t/2 + s_{q_t} \delta_t; r) \bigg) \tilde{\vec{\mathcal{U}}}_0 + \vec{E}_{\text{trunc}}^{(3)},
\end{equation}
where $s_{q_t}$ are the roots of the $Q_t$th-order Legendre polynomial $P_{Q_t}(s)$ and $w_{q_t}= \delta_t/[(1 - s_{q_t}^2) {P_{Q_t}'}^2(s_{q_t}))$ are the associated Gaussian weights.
% the Hamiltonian is assumed to be constant within each step $[t_l, t_l+\delta_t]$.
% We adopt a representative value at time $t_l+\xi_l\delta_t=(l+\xi_l)t/N_t$, where $t_l=l\delta_t$ and $\xi_l\in(0,1)$.
% This procedure leads to the approximation
% \begin{equation}\label{eq:time_evo_opt_0}
%     \vec{U}_r(t) \tilde{\vec{\mathcal{U}}}_0
%     = \prod_{l=0}^{N_t-1} \ee^{\ii\tilde{\vec{B}}((l+\xi_l)\delta_t)\delta_t} \tilde{\vec{\mathcal{U}}}_0
%     + \vec{E}_{\mathrm{trunc}}^{(3)}.
% \end{equation}
We now estimate the upper bound of the error term $\vec{E}_{\mathrm{trunc}}^{(3)}$.
Within each time step, the local error is bounded by~\cite{Hairer2006_Geometric}
\begin{equation}\label{eq:time_discrete_error}
    \bigg\|\mathcal{T}\exp\bigg(\ii\int_{n_t\delta_t}^{(n_t+1)\delta_t} \tilde{\vec{B}}(s)\dif s\bigg) \tilde{\vec{\mathcal{U}}}_0 -  \exp\bigg( \ii \sum_{q_t=0}^{Q_t-1} w_{q_t} \tilde{\tensor{B}}((2n_t+1)\delta_t/2 + s_{q_t} \delta_t; r) \bigg) \tilde{\vec{\mathcal{U}}}_0 \bigg\|_2
    \le C_B \delta_t^{2Q_t + 1} \norm{\tilde{\vec{\mathcal{U}}}_0}_2,
\end{equation}
where the constant $C_B$ depends on the properties of $\tilde{\tensor{B}}$, specifically on the upper bounds of the norms of $\tilde{\tensor{B}}(s)$ and its higher-order time derivatives within the interval $[n_t\delta_t, (n_t+1)\delta_t]$, but is independent of $\delta_t$.
Summing these local errors over all $I_t$ time steps via a telescoping series and applying the triangle inequality provides the global error bound
\begin{equation}
    \norm{\vec{E}_{\text{trunc}}^{(3)}}_2
    \le \sum_{n_t=0}^{I_t - 1} C_B \delta_t^{2Q_t + 1} \norm{\tilde{\vec{\mathcal{U}}}_0}_2
    = C_B t \delta_t^{2Q_t} \norm{\tilde{\vec{\mathcal{U}}}_0}_2.
\end{equation}
For notational simplicity, we collapse the double product over time steps and quadrature points into a single product $\prod_{l=0}^{N_t-1}$ with $N_t=I_tQ_t$, which yields
\begin{equation}\label{eq:time_evo_opt_11}
    \vec{U}_{r_j}(t) = \prod_{l=0}^{N_t - 1}\exp\left[\ii w_l (\tensor{H}_l + r_j\tilde{\tensor{L}}_l\right] + O(t\delta_t^{2Q_t}),
\end{equation}
where $\tensor{H}_l = \tensor{H}((2n_t+1)\delta_t/2 + s_{q_t}\delta_t)$, $\tilde{\tensor{L}}_l = \tilde{\tensor{L}}((2n_t+1)\delta_t/2 + s_{q_t}\delta_t)$ and $w_l=w_{q_t}$ with $l = q_t + n_t Q_t$.

Since $\vec{H}$ and $\tilde{\vec{L}}$ do not commute, we further employ a $p$-th order Trotter-Suzuki (TS) decomposition~\cite{Suzuki1990_Fractal, Suzuki1991_General}
\begin{equation}\label{eq:pth_TS_decom}
    \ee^{\ii(\vec{A}+\vec{B})\dt} = \prod_{k=1}^{N_p} \ee^{\ii \alpha_k \vec{A} \delta_t} \ee^{\ii \beta_k \vec{B} \delta_t} + O(\delta_t^{p+1}),
\end{equation}
where $N_p\in\mathbb{N}$ is the number of operators in the product expansion, and the real coefficients $\{\alpha_k\}_{k=1}^{N_p}$ and $\{\beta_k\}_{k=1}^{N_p}$ are chosen to cancel error terms up to order $p$.
Applying Eq.~\eqref{eq:pth_TS_decom} to $\vec{U}_{r_j}(t)$, we obtain
\begin{equation}\label{eq:pth_TS_decom_app}
    \ee^{\ii w_l (\tensor{H}_l + r_j\tilde{\tensor{L}}_l)}
    = \prod_{k=1}^{N_p} \ee^{\ii \alpha_k w_l \vec{H}_l} \ee^{\ii r_j\beta_k w_l\tilde{\vec{L}}_l} + O(\delta_t^{p+1}).
\end{equation}
Substituting Eq.~\eqref{eq:pth_TS_decom_app} into Eq.~\eqref{eq:time_evo_opt_11} yields
\begin{equation}\label{eq:pth_TS_decom_app_1}
    \begin{aligned}
        \vec{U}_{r_j}(t)
        &= \prod_{l=0}^{N_t-1} \bigg(\prod_{k=1}^{N_p} \ee^{\ii \alpha_k w_l \vec{H}_l} \ee^{\ii r_j\beta_k w_l \tilde{\vec{L}}_l} + O(\delta_t^{p+1}) \bigg)
        + O(t \delta_t^{2Q_t})
        \\
        &= \prod_{l=0}^{N_t-1} \prod_{k=1}^{N_p} \ee^{\ii \alpha_k w_l \vec{H}_l} \ee^{\ii r_j\beta_k w_l \tilde{\vec{L}}_l} + O(\max\{t\delta_t^p Q_t, t\delta_t^{2Q_t}\}).
    \end{aligned}
\end{equation}
Substituting Eq.~\eqref{eq:pth_TS_decom_app_1} into Eq.~\eqref{eq:Ut_0} yields the approximation
\begin{equation}\label{eq:pth_TS_decom_app_2}
    \tilde{\vec{\mathcal{U}}}(t) = \sum_{j=0}^{M-1} c_j \prod_{l=0}^{N_t-1} \prod_{k=1}^{N_p} \ee^{\ii \alpha_k w_l \vec{H}_l} \ee^{\ii r_j\beta_k w_l\tilde{\vec{L}_l} }\tilde{\vec{\mathcal{U}}}_0 + O(\ee^{-R^\beta}, R(\ee ht/2)^{2Q}, \max\{t\delta_t^p Q_t, t\delta_t^{2Q_t}\}).
\end{equation}
To achieve a target precision $\varepsilon$, the parameters are chosen to ensure each error term is of order $O(\varepsilon)$.
For the spatial discretization, we set $h = 1/(\ee t)$.
The error constraints $\ee^{-R^\beta} = O(\varepsilon)$ and $R(\ee ht/2)^{2Q} = O(\varepsilon)$ then require the scaling relations
\begin{equation}
    R = O(\log^{1/\beta}(1/\varepsilon)), \quad
    Q = O(\log(1/\varepsilon)).
\end{equation}
Consequently, the total number of spatial quadrature points becomes $M = 2RQ/h = O(t \log^{1+1/\beta}(1/\varepsilon))$.
For the temporal discretization, using $I_t = 2t$ time intervals and a $p$th-order quadrature with $p=2Q_t$, the constraint $t\delta_t^{2Q_t} = O(\varepsilon)$ implies
\begin{equation}
    Q_t = O(\log(1/\varepsilon)), \quad
    N_t = Q_tI_t = O(t \log(1/\varepsilon)).
\end{equation}

% According to Eq.~\eqref{eq:pth_TS_decom_app_1}, the first-order TS decomposition introduces a dominant error of $O(t^2/N_t)$, whereas the error associated with a second-order or higher decomposition, $O(t^3/N_t^2)$, is commensurate with the temporal discretization error in Eq.~\eqref{eq:time_discrete_error}.
% Based on this scaling, we employ a second-order TS decomposition in the subsequent analysis, whereupon Eq.~\eqref{eq:pth_TS_decom_app_2} takes the form
% \begin{equation}\label{eq:LCU_evolve_eq}
%     \tilde{\vec{\mathcal{U}}}(t) = \sum_{j=0}^{M-1} c_j \prod_{l=0}^{N_t-1} \ee^{\ii \vec{H}((l+\frac12)\delta_t) \delta_t/2} \ee^{\ii r_j \tilde{\vec{L}}((l+\frac12)\delta_t) \delta_t} \ee^{\ii \vec{H}((l+\frac12)\delta_t) \delta_t/2} \tilde{\vec{\mathcal{U}}}_0 + O\bigg(\frac{R^3}{M^2}, \ee^{-R^\beta}, \frac{Mt^3}{N_t^2} \bigg).
% \end{equation}
\section{End-to-end quantum algorithm for simulating rapidly distorted turbulence}\label{sec:quantum_algorithm}
We propose an end-to-end quantum algorithm for simulating turbulence via RDT, encompassing state preparation, time evolution, and statistics extraction through measurements.
Each step is implemented with a quantum circuit of polynomial depth, resulting in a time complexity that scales polynomially with the number of qubits.
This algorithm thereby achieves an end-to-end quantum speedup over its classical counterpart.

We first detail the quantum encoding of the flow field, defined on a spatial grid of $N=N_x N_y N_z$ total points, where $N_\alpha$ for $\alpha \in \{x,y,z\}$ is the number of points in each direction.
The velocity field $\tilde{\vec{\mathcal{U}}}$ is encoded using $n=\lceil\log_2 N\rceil+2$ qubits.
For an initial field normalized such that $\norm{\tilde{\vec{\mathcal{U}}}_0}_2=1$, the quantum state is
\begin{equation}\label{eq:encode}
    \ket{\tilde{\mathcal{U}}(t)}
    = \sum_{m=0}^{N-1} \sum_{i=0}^2 \hat{u}_i(\vec{\kappa}_m, t)\ket{m, i},
\end{equation}
where $n_N=\lceil\log_2 N\rceil$ qubits constitute the grid index register $\ket{m}$ and $n_d=2$ qubits form the component index register $\ket{i}$.
The component indices $i \in \{0,1,2\}$ are encoded as the computational basis states $\ket{00}$, $\ket{01}$, and $\ket{10}$, respectively, thereby utilizing three of the four available basis states of the component register.
A given grid index $m$ uniquely determines the spatial indices $(m_x,m_y,m_z)$ via the explicit mapping $m_x=(m \mod N_x)$, $m_y=(\lfloor m/N_x\rfloor \mod N_y)$, and $m_z= \lfloor m/(N_xN_y)\rfloor$, with the inverse mapping being $m=m_x+N_x m_y + N_x N_y m_z$.

\subsection{Turbulent state preparation}
We use $n_c = \lceil \log_2 M \rceil$ ancillary qubits to encode the LCU coefficients $c_j$ in Eq.~\eqref{eq:pth_TS_decom_app_2}.
Starting from the ground state $\ket{0}_c^{\otimes n_c}\otimes \ket{0}^{\otimes n}$, we apply the unitary transformations $U_{\mathrm{prep}}$ and $U_{\mathrm{coef}}$ to prepare the initial condition and the coefficients, respectively, which yields the state
\begin{equation}\label{eq:state_preparation_target}
    (U_{\mathrm{coef}} \otimes U_{\mathrm{prep}}) \ket{0}_c^{\otimes n_c}\otimes \ket{0}^{\otimes n}
    = \frac{1}{\sqrt{\norm{\vec{c}}_1}} \sum_{j=0}^{M-1} \sqrt{c_j} \ket{j}_c \otimes \ket{\tilde{\mathcal{U}}(0)}.
\end{equation}
Because the function $\sqrt{c_j}$ and its derivatives are smooth and structured, $U_{\mathrm{coef}}$ can be implemented with precision $\varepsilon$ at a negligible cost of $O(\log M\cdot\log(1/\varepsilon))$ using methods such as quantum polynomial approximation or hierarchical construction~\cite{Soklakov2006_Efficient}.

The cost of $U_{\mathrm{prep}}$ depends on the structure of the initial state $\ket{\tilde{\mathcal{U}}(0)}$.
In RDT, the normalized initial field is typically chosen as
\begin{equation}\label{eq:IC_vel}
    \hat{\vec{u}}(\vec{k}, t=0) = \frac{1}{\norm{E_0}_1^{1/2}}\sqrt{E_0(|\vec{k}|)} \ee^{\ii\phi_{\vec{k}}}\vec{k}_{\perp},
\end{equation}
where $\phi_{\vec{k}}$ is a random phase uniformly distributed in $[0,2\pi]$, $\vec{k}_{\perp}$ is a random unit vector satisfying $\vec{k}\cdot\vec{k}_{\perp}=0$, and $E_0(|\vec{k}|)$ is a specified energy spectrum.
A common choice for the initial energy spectrum is the Passot-Pouquet form $E_0(|\vec{k}|)=|\vec{k}|^4 \ee^{-2(|\vec{k}|/k_p)^2}$, where $k_p$ is the peak wavenumber~\cite{Passot1987_Numerical}.
Another is the model spectrum $E_0(|\vec{k}|) = k^{-5/3} f_L(kL) f_\eta(k_\eta)$, where $f_L(kL) = (k L / \sqrt{(kL)^2 + c_L})^{5/3 + p_0}$, $f_\eta(k\eta) = \exp\{-\beta[((k\eta)^4 + c_\eta^4)^{1/4} - c_\eta] \}$, and the parameters $L$, $\eta$, $c_L$, $c_\eta$, and $\beta$ are adjustable~\cite{Pope2000_Turbulent}.
If the pseudo-random phase $\phi_{\vec{k}}$ and the components of the random unit vector $\vec{k}_{\perp}$ are generated by an efficiently computable classical function of the wavenumber index $m$, the corresponding quantum oracles for phase and wavevector loading possess a polynomial gate complexity.
This ensure the entire initial quantum state
\begin{equation}\label{eq:ket_U0_target}
    \ket{\tilde{\mathcal{U}}(0)}
    = \sum_{m=0}^{N-1} \sum_{i=0}^2 A_m \ee^{\ii\phi_{m, i}} k_{\perp, m, i} \ket{m, i}
    = \sum_{m=0}^{N-1} A_m\ket{m} \otimes \sum_{i=0}^2 \ee^{\ii\phi_{m, i}} k_{\perp, m, i}\ket{i}
\end{equation}
is efficiently preparable with polynomial quantum resources because $E_0(|\vec{k}|)$ is a smooth, integrable function.
Here, $A_m = \sqrt{E_0(|\vec{k}_m|)/\|E_0\|_1}$ is the normalized real amplitude, and $k_{\perp, m, i}$ is the $i$-th Cartesian component of $\vec{k}_{\perp}(m)$.

We decompose the preparation of this multipartite state in Eq.~\eqref{eq:ket_U0_target} into three steps: loading the real amplitudes $A_m$ associated with the grid index $m$, loading the components of $\vec{k}_\perp$ conditioned on $m$, and finally applying the corresponding complex phases $\ee^{\ii\phi_{m, i}}$.
This procedure begins with the ground state $\ket{0}^{\otimes n_N}\otimes \ket{00}$ and incrementally constructs the target state through the sequential application of three unitary operators, $U_A$, $U_k$, and $U_\phi$.

\subsubsection*{Step 1: Amplitude loading}
The first step is amplitude loading, wherein the unitary operator $U_A$ is applied to the grid register to prepare the correct superposition of probability amplitudes.
The function of this operation is
\begin{equation}\label{eq:Amplitude_loading_state}
    U_A\otimes I^{\otimes 2}: \ket{0}^{\otimes n_N}\otimes \ket{00} \mapsto \bigg(\sum_{m=0}^{N-1} A_m\ket{m} \bigg) \otimes \ket{00}.
\end{equation}
Since the amplitude function $A_{m}$ is real, smooth, efficiently computable, and unimodal with respect to the index $m$, $U_A$ can be efficiently implemented via a standard quantum state preparation algorithm based on controlled rotations.
The detailed implementation is as follows.

First, the operator $U_A$ is decomposed into a product of $n_N$ unitary operators as
\begin{equation}
    U_A = \prod_{j=1}^{n_N} U_j,
\end{equation}
where each $U_j$ acts on the $j$-th qubit, controlled by the preceding $j-1$ qubits.
For a binary prefix $b=(b_j\cdots b_2b_1)_2$ of length $j$, corresponding to the binary representation of an index $m=(b_{n_N}\cdots b_2b_1)_2$, we define
\begin{equation}
    P_b := \sum_{m \in S_b} A_m^2,
\end{equation}
which is the sum of probabilities for all states whose indices share the prefix $b$, with the index set being $S_b = \{m \mid \lfloor m/2^{n_N-j} \rfloor = b\}$.
For an empty prefix ($j=0$), this definition implies $P_b=\sum_{m=0}^{N-1} A_m^2=1$.

\begin{figure}[t!]
    \centering
    \includegraphics{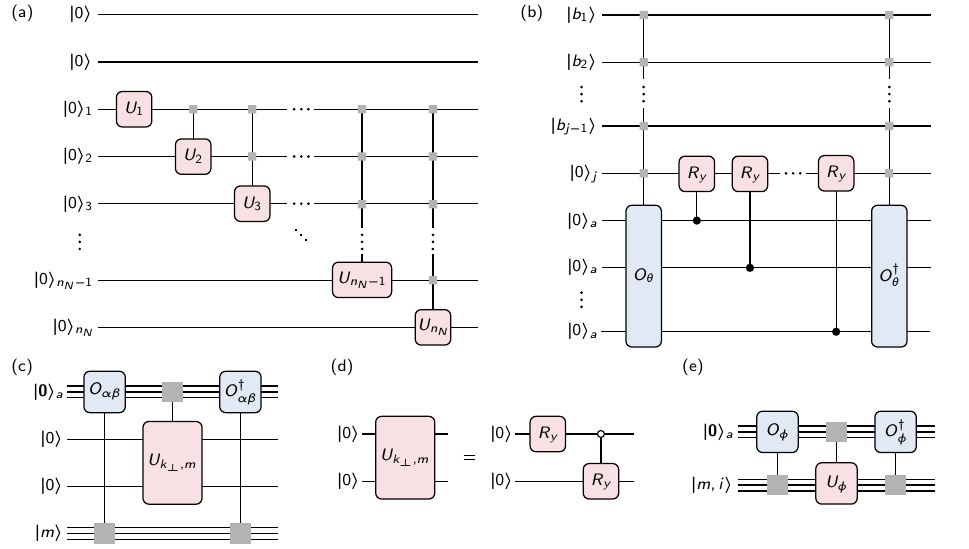}
    \caption{Quantum circuits for state preparation.
    (a) Decomposition of the operator $U_A$ for amplitude loading, as defined in Eq.~\eqref{eq:Amplitude_loading_state}.
    (b) Circuit implementation of the unitary $U_j$ from Eq.~\eqref{eq:U_j_amplitude}, corresponding to the sequence in Eq.~\eqref{eq:amplitude_sequence}.
    (c) Circuit implementation of the unitary $U_{k_\perp}$ for wavevector loading, corresponding to the sequence in Eq.~\eqref{eq:wavevector_loading_state}.
    (d) Decomposition of each constituent operator $U_{k_\perp, m}$.
    (e) Circuit implementation of the unitary $U_{\phi}$ for phase loading, corresponding to the sequence in Eq.~\eqref{eq:phase_sequence}.
    The grey boxes denote control qubits or registers without specifying the control state ($\ket{0}$ or $\ket{1}$).
    }
    \label{fig:circuit_state_preparation}
\end{figure}

Then we apply $U_1=I^{\otimes (n_N-1)} \otimes R_y(\theta_1)$, where the angle $\theta_1=2\arccos\sqrt{P_0}$ with $P_0=\sum_{m=0}^{N/2-1} A_m^2$ prepares the first qubit.
The function of this operation is
\begin{equation}
    U_1: \ket{0}^{\otimes n_N} \mapsto \ket{0}^{\otimes (n_N-1)} \otimes (\cos(\theta_1/2)\ket{0} + \sin(\theta_1/2)\ket{1})
    = \ket{0}^{\otimes (n_N-1)} \otimes (\sqrt{P_0}\ket{0} + \sqrt{P_1}\ket{1}).
\end{equation}
Iteratively, in the $j$-th step ($j=2,3,\cdots,n_N$), we apply the multi-controlled rotation gate
\begin{equation}\label{eq:U_j_amplitude}
    U_j = I^{\otimes(n_N-j)} \otimes \sum_{b=0}^{2^{j-1}-1} R_y(\theta_{j|b}) \otimes \ket{b}\bra{b}.
\end{equation}
Here, the rotation angle $\theta_{j|b}=2\arccos\sqrt{P_{0b}/P_b}$ is conditioned on the state of the preceding $j-1$ qubits, $\ket{b}=\ket{b_{j-1}\cdots b_2b_1}$, with $P_{0b}$ denoting the probability associated with the length-$j$ prefix $\ket{0b}$.
Assuming that after $j-1$ steps the system is in the state
\begin{equation}
    U_{j-1}\cdots U_1: \ket{0}^{\otimes n_N} \mapsto \ket{0}^{\otimes (n_N-j+1)} \otimes \sum_{b=0}^{2^{j-1}-1} \sqrt{P_b}\ket{b},
\end{equation}
the subsequent application of $U_j$ yields
\begin{align}
    U_j \left( \ket{0}^{\otimes (n_N-j+1)} \otimes \sum_{b=0}^{2^{j-1}-1} \sqrt{P_b}\ket{b} \right)
    &= \left( I^{\otimes(n_N-j)} \otimes \sum_{b=0}^{2^{j-1}-1} R_y(\theta_{j|b}) \otimes \ket{b}\bra{b} \right) \left(\ket{0}^{\otimes (n_N-j+1)} \otimes \sum_{b'=0}^{2^{j-1}-1} \sqrt{P_{b'}}\ket{b'} \right) \nonumber
    \\ 
    &= \ket{0}^{\otimes (n_N-j)} \otimes \sum_{b=0}^{2^{j-1}-1} \sqrt{P_b} \left( R_y(\theta_{j|b})\ket{0} \right) \otimes \ket{b} \nonumber
    \\ 
    &= \ket{0}^{\otimes (n_N-j)} \otimes \sum_{b=0}^{2^{j-1}-1} \sqrt{P_b} \left(\sqrt{P_{0b}/P_b}\ket{0} + \sqrt{P_{1b}/P_b}\ket{1}\right) \otimes \ket{b} \nonumber
    \\
    &= \ket{0}^{\otimes (n_N-j)} \otimes \sum_{b=0}^{2^{j-1}-1} \left(\sqrt{P_{0b}}\ket{0} + \sqrt{P_{1b}}\ket{1}\right) \otimes \ket{b} \nonumber
    \\
    &= \ket{0}^{\otimes (n_N-j)} \otimes \sum_{b'=0}^{2^j-1} \sqrt{P_{b'}}\ket{b'}.
\end{align}
By induction, after $n_N$ steps, the total operation results in the state
\begin{equation}\label{eq:U_A}
    U_A: \ket{0}^{\otimes n_N} \mapsto \sum_{b=0}^{2^{n_N}-1} \sqrt{P_b}\ket{b}.
\end{equation}
At the final step $j=n_N$, the prefix $b$ corresponds to the index $m$, such that $P_b=A_m^2$.
Equation~\eqref{eq:U_A} thus represents the desired prepared state, with quantum circuit in Fig.~\ref{fig:circuit_state_preparation}(a).

To implement the operator $U_j$, a quantum circuit must compute the angle $\theta_{j|b}$ conditioned on the control state $\ket{b}$.
Since $A_m$ is efficiently computable, the summation for $P_b$ and $P_{0b}$, followed by the arithmetic operations for $\theta_{j|b}$, can all be performed efficiently on a quantum computer.
The specific implementation of each $U_j$ in Fig.~\ref{fig:circuit_state_preparation}(b) employs an ancillary register, initialized to $\ket{\vec{0}}_a$, and follows the sequence
\begin{equation}\label{eq:amplitude_sequence}
    \ket{\vec{0}}_a\ket{0}_j\ket{b} 
    \xrightarrow{O_\theta} \ket{\theta_{j|b}}_a\ket{0}_j\ket{b} 
    \xrightarrow{C\text{-}R_y} \ket{\theta_{j|b}}_a R_y(\theta_{j|b})\ket{0}_j \ket{b}
    \xrightarrow{O_\theta^\dagger} \ket{\vec{0}}_a R_y(\theta_{j|b})\ket{0}_j \ket{b},
\end{equation}
where $\ket{b}$ is the state of the first $j-1$ qubits, $\ket{0}_j$ is the $j$-th qubit state, $O_\theta$ is an angle-computation oracle, and $C\text{-}R_y$ is a rotation controlled by the ancillary register.
The properties of $A_m$ allow the oracle $O_\theta$ to be constructed with a circuit of polynomial size.
Consequently, the circuit depth of the entire amplitude loading process is $O(\poly(\log N))$.

\subsubsection*{Step 2: Wavevector loading}
Next, the grid-index-dependent amplitudes of the wavenumber components are prepared.
This is achieved by applying a controlled unitary operator
\begin{equation}\label{eq:U_k}
    U_{k_\perp} = \sum_{m=0}^{N-1} \ket{m}\bra{m} \otimes U_{k_\perp, m}
\end{equation}
to entangle the component index register with the grid index register.
Each operator $U_{k_\perp,m}$ acts on the component register such that
\begin{equation}
    U_{k_\perp,m}\ket{00} = \sum_{i=0}^2 k_{\perp, m, i} \ket{i}.
\end{equation}
This transformation is decomposed into controlled rotations as $U_{k_\perp,m}=C_0^{(1)}(R_y^{(2)}(\alpha_m))R^{(1)}(\beta_m)$, where $R^{(1)}(\beta_m)$ acts on the first qubit, and $C_0^{(1)}(R_y^{(2)}(\alpha_m))$ is a rotation on the second qubit, conditional on the first qubit being in the state $\ket{0}$, as shown in Fig.~\ref{fig:circuit_state_preparation}(d).
The rotation angles are given by $\alpha_m=2\arctan2(k_{\perp,m,1}, k_{\perp,m,0})$ and $\beta_m=2\arcsin(k_{\perp,m,2})$.
Applying $U_{k_\perp}$ to the state given by Eq.~\eqref{eq:Amplitude_loading_state} yields
\begin{equation}\label{eq:wavevector_loading_state}
    U_{k_\perp}: \bigg(\sum_{m=0}^{N-1} A_m\ket{m} \bigg) \otimes \ket{00} \mapsto \sum_{m=0}^{N-1} A_m\ket{m} \otimes \sum_{i=0}^2 k_{\perp, m, i}\ket{i}.
\end{equation}

The implementation of $U_{k_\perp}$ in Fig.~\ref{fig:circuit_state_preparation}(c) employs an ancillary register, analogous to the previous step, and proceeds as
\begin{equation}
    \ket{m}\ket{00}\ket{\vec{0}}_a \xrightarrow{O_{\alpha\beta}} \ket{m}\ket{00}\ket{\alpha_m,\beta_m}_a
    \xrightarrow{U_{k_\perp}} \ket{m} (U_{k_\perp,m}\ket{00}) \ket{\alpha_m, \beta_m}_a
    \xrightarrow{O_{\alpha\beta}^\dagger} \ket{m} (U_{k_\perp,m}\ket{00}) \ket{\vec{0}}_a.
\end{equation}
Here, the oracle $O_{\alpha\beta}$ computes the angles $\alpha_m$ and $\beta_m$ into the ancillary register, which then controls the required rotations.
Efficient implementation requires that $\alpha_m$ and $\beta_m$ be efficiently computable, which is true if the vector $\vec{k}_\perp$ is generated by an efficiently computable pseudo-random function of $m$, such as one based on a Gram-Schmidt process.
Under this ansatz, the gate complexity for $U_{k_\perp}$ is $O(\poly(\log N))$.

\subsubsection*{Step 3: Phase loading}
Finally, a diagonal unitary operator
\begin{equation}\label{eq:U_phi}
    U_\phi = \sum_{m=0}^{N-1}\sum_{i=0}^2 \ee^{\ii\phi_{m,i}}\ket{m,i}\bra{m,i}
\end{equation}
encodes the phase $\ee^{\ii\phi_{m,i}}$ onto each computational basis state $\ket{m,i}$.
Applying this operator to the state from the preceding step in Eq.~\eqref{eq:wavevector_loading_state} yields
\begin{equation}
    U_\phi:
    \sum_{m=0}^{N-1} A_m\ket{m} \otimes \sum_{i=0}^2 k_{\perp, m, i}\ket{i} \mapsto \sum_{m=0}^{N-1}\sum_{i=0}^2 A_m k_{\perp,m,i}\ee^{\ii\phi_{m,i}}\ket{m,i},
\end{equation}
thereby completing the preparation of the initial state in Eq.~\eqref{eq:ket_U0_target}.

Provided that $\phi_{m,i}$ is efficiently computable by a pseudo-random number generator, the diagonal operator $U_\phi$ is implemented with a circuit depth of $O(\poly(\log N))$ via the phase kickback technique~\cite{Cleve1998_Quantum}.
This implementation, shown in Fig.~\ref{fig:circuit_state_preparation}(e), requires an ancillary register and proceeds as
\begin{equation}\label{eq:phase_sequence}
    \ket{m,i}\ket{\vec{0}}_a
    \xrightarrow{O_\phi} \ket{m,i}\ket{\phi_{m,i}}_a
    \xrightarrow{U_\phi} \ee^{\ii\phi_{m,i}}\ket{m,i}\ket{\phi_{m,i}}_a
    \xrightarrow{O_\phi^\dagger} \ee^{\ii\phi_{m,i}}\ket{m,i}\ket{\vec{0}}_a,
\end{equation}
where the oracle $O_\phi$ computes the phase $\phi_{m,i}$ conditioned on the state $\ket{m,i}$ and stores it as a binary fixed-point number in the ancillary register.

The complete quantum circuit for initial state preparation is thus given by the unitary operator
\begin{equation}
    U_{\mathrm{prep}} = U_\phi U_{k_\perp} (U_A\otimes I^{\otimes 2}).
\end{equation}
Application of this operator to the ground state $\ket{0}^{\otimes n}$ produces the initial state specified in Eq.~\eqref{eq:ket_U0_target}, which encodes the RDT initial conditions.
Since the parameters $A_m$, $k_{\perp,m,i}$, and $\phi_{m,i}$ arise from smooth functions and efficient pseudo-random number generation, each constituent operator, $U_A$, $U_{k_\perp}$, and $U_\phi$, is implementable by a quantum circuit of polynomial depth.
The entire preparation process is therefore efficient.

\subsection{Time evolution}
We then apply the operator
\begin{equation}
    U_{\mathrm{select}} = \sum_{j=0}^{M-1} \ket{j}_c\bra{j}_c\otimes U_{r_j}(t)
\end{equation}
to the prepared state in Eq.~\eqref{eq:state_preparation_target}, where $U_{r_j}(t)$ is the approximate evolution given by Eq.~\eqref{eq:pth_TS_decom_app_1}.
The application of this operator yields
\begin{equation}
    U_{\mathrm{select}}: 
    \frac{1}{\sqrt{\norm{\vec{c}}_1}} \sum_{j=0}^{M-1} \sqrt{c_j} \ket{j}_c \otimes \ket{\tilde{\mathcal{U}}(0)}
    \mapsto
    \frac{1}{\sqrt{\norm{\vec{c}}_1}} \sum_{j=0}^{M-1} \sqrt{c_j} \ket{j}_c \otimes U_{r_j}(t)\ket{\tilde{\mathcal{U}}(0)},
\end{equation}
thereby entangling each basis state $\ket{j}_c$ of the ancillary register with the corresponding evolved state $U_{r_j}(t)\ket{\tilde{\mathcal{U}}(0)}$ of the main system.

According to Eq.~\eqref{eq:pth_TS_decom_app_1}, $U_{\mathrm{select}}$ is approximated by the decomposition
\begin{equation}\label{eq:U_select_Np}
    U_{\mathrm{select}}
    = \prod_{l=0}^{N_t-1} \prod_{k=1}^{N_p} \bigg( \sum_{j=0}^{M-1} \ket{j}_c\bra{j}_c \otimes V_{j,k,l}\bigg)
\end{equation}
with
\begin{equation}
    V_{j,k,l} = \ee^{\ii \alpha_k w_l \vec{H}_l} \ee^{\ii r_j\beta_k w_l \tilde{\vec{L}}_l}.
\end{equation}
Within the exponents of $V_{j,k,l}$, all parameters are held constant except for $r_j$.
Consequently, the exponent of the first term remains constant, while that of the second term exhibits a linear dependence on $r_j$.
This linear structure enables the parallel implementation of these operators using quantum arithmetic circuits, which necessitates an additional work register.
The resulting total complexity is therefore $O(N_tN_p\poly(\log M))=O(N_p t \log(1/\varepsilon))$.
The large coefficients $\alpha_k$ and $\beta_k$ inherent in high-order TS decompositions can introduce significant numerical errors.
To balance precision and computational cost, we therefore employ the second-order TS decomposition, for which $N_p=2$.
The operator $U_{\mathrm{select}}$ in Eq.~\eqref{eq:U_select_Np} consequently reduces to
\begin{equation}\label{eq:U_select}
    U_{\mathrm{select}}
    = \prod_{l=0}^{N_t-1} \bigg( \sum_{j=0}^{M-1} \ket{j}_c\bra{j}_c \otimes V_{j,l}\bigg)
\end{equation}
with
\begin{equation}\label{eq:opt_each}
    V_{j,l} = \ee^{\ii w_l \vec{H}_l/2} \ee^{\ii r_j w_l \tilde{\vec{L}}_l} \ee^{\ii w_l \vec{H}_l/2}.
\end{equation}

Then, we describe the implementation of Eq.~\eqref{eq:opt_each} at the quantum gate level.
As the three exponential factors in Eq.~\eqref{eq:opt_each} are structurally identical, we illustrate the method with the central term, which we denote for brevity as $\ee^{\ii r_j w_l \tilde{\vec{L}}_l}$.
Substituting Eq.~\eqref{eq:def_L}, this operator is expanded as
\begin{equation}\label{eq:imple_0}
    \ee^{\ii r_j w_l \tilde{\vec{L}}_l}
    = \exp\bigg[\ii r_j w_l \bigg(\sum_{m=0}^{N-1} \ket{m}\bra{m}\otimes \vec{L}_{d}^{(l)}(\vec{\kappa}_m) - cI^{\otimes n}\bigg) \bigg],
\end{equation}
where $\vec{L}_{d}^{(l)}(\vec{\kappa}_m)$ is the block-diagonal matrix
\begin{equation}
    \vec{L}_{d}^{(l)}(\vec{\kappa}_m)
    = \bigg(\frac{\vec{\kappa}_m(l)\vec{\kappa}_m\T(l)\vec{A}(l) + \vec{A}\T(l)\vec{\kappa}_m(l)\vec{\kappa}_m\T(l)}{|\vec{\kappa}_m(l)|^2} - \frac{\vec{A}(l) + \vec{A}\T(l)}{2} \bigg) \oplus 0
    := \begin{bmatrix}
        l_{11} & l_{12} & l_{13} & 0 \\
        l_{12} & l_{22} & l_{23} & 0 \\
        l_{13} & l_{23} & l_{33} & 0 \\
        0 & 0 & 0 & 0
    \end{bmatrix}.
\end{equation}
The elements of the non-zero block are given by
\begin{equation}\label{eq:h_ij}
    l_{ij}(m, l) = \frac{\kappa_i(m, l)\kappa_k(m, l) a_{kj}(l) + \kappa_j(m, l)\kappa_k(m, l) a_{ki}(l)}{\kappa_k(m, l) \kappa_k(m, l)} - \frac{a_{ij}(l) + a_{ji}(l)}{2}, \quad i,j\in\{1,2,3\}.
\end{equation}

We decompose the matrix $\vec{L}_{d}^{(l)}(\vec{\kappa}_m)$ into a sum of its diagonal and off-diagonal components as
\begin{equation}\label{eq:decompose_Ld}
    \vec{L}_{d}^{(l)}(\vec{\kappa}_m)
    = \vec{L}_D + \vec{L}_{12} + \vec{L}_{13} + \vec{L}_{23},
\end{equation}
where the four terms
\begin{equation}
    \vec{L}_D = \begin{bmatrix}
        l_{11} & 0 & 0 & 0 \\
        0 & l_{22} & 0 & 0 \\
        0 & 0 & l_{33} & 0 \\
        0 & 0 & 0 & 0
    \end{bmatrix}, \
    \vec{L}_{12} = \begin{bmatrix}
        0 & l_{12} & 0 & 0 \\
        l_{12} & 0 & 0 & 0 \\
        0 & 0 & 0 & 0 \\
        0 & 0 & 0 & 0
    \end{bmatrix}, \
    \vec{L}_{13} = \begin{bmatrix}
        0 & 0 & l_{13} & 0 \\
        0 & 0 & 0 & 0 \\
        l_{13} & 0 & 0 & 0 \\
        0 & 0 & 0 & 0
    \end{bmatrix}, \
    \vec{L}_{23} = \begin{bmatrix}
        0 & 0 & 0 & 0 \\
        0 & 0 & l_{23} & 0 \\
        0 & l_{23} & 0 & 0 \\
        0 & 0 & 0 & 0
    \end{bmatrix}
\end{equation}
represent the distortion in various spatial directions caused by the mean shear flow.
Substituting Eq.~\eqref{eq:decompose_Ld} into Eq. \eqref{eq:imple_0} yields
\begin{equation}\label{eq:imple_1}
    \ee^{\ii r_j w_l \tilde{\vec{L}}_l}
    = \bigg(\sum_{m=0}^{N-1} \ket{m}\bra{m} \otimes \ee^{\ii r_j w_l (\vec{L}_D + \vec{L}_{12} + \vec{L}_{13} + \vec{L}_{23})}\bigg)\ee^{-\ii cr_j w_l}.
\end{equation}
To match the temporal precision, we apply a second-order TS decomposition
\begin{equation}\label{eq:evolution_element}
    \ee^{\ii r_j w_l (\vec{L}_D + \vec{L}_{12} + \vec{L}_{13} + \vec{L}_{23})}
    = \ee^{\ii r_jw_l \vec{L}_{23}/2} \ee^{\ii r_jw_l \vec{L}_{13}/2} \ee^{\ii r_jw_l \vec{L}_{12}/2} \ee^{\ii r_jw_l \vec{L}_{D}} \ee^{\ii r_jw_l \vec{L}_{12}/2} \ee^{\ii r_jw_l \vec{L}_{13}/2} \ee^{\ii r_jw_l \vec{L}_{23}/2} + O(\delta_t^3)
\end{equation}
to the operator in Eq. \eqref{eq:imple_1}.

We proceed by decomposing each term in Eq.~\eqref{eq:evolution_element} into a sequence of elementary quantum gates.
First, the exponentiation of the diagonal term $\vec{L}_{D}$ is direct, yielding the operator $\ee^{\ii r_jw_l \vec{L}_{D}} = \diag(\ee^{\ii r_jw_l l_{11}}, \ee^{\ii r_jw_l l_{22}}, \ee^{\ii r_jw_l l_{33}}, 1)$.
This transformation is realized by applying a phase of $\ee^{\ii r_jw_l(l_{22}-l_{11})}$ to the basis state $\ket{2}$ and a phase of $\ee^{\ii r_jw_l(l_{33}-l_{11})}$ to the basis state $\ket{3}$, together with an overall global phase of $\ee^{\ii r_jw_l l_{11}}$.

Subsequently, the operator $\ee^{\ii r_jw_l \vec{L}_{23}/2}$ acts non-trivially only within the two-dimensional subspace spanned by $\{\ket{2}, \ket{3}\}$, where it is proportional to the Pauli-$X$ operator.
This unitary operator
\begin{equation}
    \ee^{\ii r_jw_l \vec{L}_{23}/2} = \begin{bmatrix}
        1 & 0 & 0 & 0 \\
        0 & \cos(l_{23}r_jw_l/2) & \ii\sin(l_{23}r_jw_l/2) & 0 \\
        0 & \ii\sin(l_{23}r_jw_l/2) & \cos(l_{23}r_jw_l/2) & 0 \\
        0 & 0 & 0 & 1
    \end{bmatrix}
\end{equation}
corresponds to a generalized $R_x$ gate acting on the $\{\ket{2}, \ket{3}\}$ subspace with a rotation angle of $\theta_{23}=l_{23}r_jw_l$.
Similarly, the operators $\ee^{\ii r_jw_l \vec{L}_{12}/2}$ and $\ee^{\ii r_jw_l \vec{L}_{13}/2}$ implement $R_x$ rotations in the $\{\ket{1},\ket{2}\}$ and $\{\ket{1},\ket{3}\}$ subspaces, with respective rotation angles of $\theta_{12}=l_{12}r_jw_l$ and $\theta_{13}=l_{13}r_jw_l$.

Continuing with $\vec{L}_{23}$ as a representative example, the operator to be implemented is
\begin{equation}\label{eq:opt_H23}
    \sum_{m=0}^{N-1} \ket{m}\bra{m} \otimes \ee^{\ii r_jw_l \vec{L}_{23}/2}
    = \sum_{m=0}^{N-1} \ket{m}\bra{m} \otimes R_x^{(2,3)}(\theta_{23}(\vec{\kappa}_m)).
\end{equation}
The rotation angle $\theta_{23}(\vec{\kappa}_m)$ is an arithmetic function of the index $m$.
This property follows from Eq.~\eqref{eq:h_ij}, combined with the fact that $\vec{\kappa}_m$ is a linear function of $m$ according to Eq.~\eqref{eq:kappa(t)}.
Consequently, this controlled rotation can be implemented efficiently via a quantum arithmetic circuit with the introduction of ancillary qubits.

To this end, we introduce an ancillary register of $n_{\theta}$ qubits, initialized to the state $\ket{0}^{\otimes n_\theta}$, where $n_\theta$ determines the precision of the angle computation.
We then apply a quantum arithmetic circuit, $O_{23}$, that performs the transformation
\begin{equation}\label{eq:O_23}
    O_{23}: \ket{\vec{\kappa}} \otimes \ket{0}_a^{\otimes n_\theta} \mapsto \ket{\vec{\kappa}} \otimes \ket{\theta_{23}(\vec{\kappa})}_a.
\end{equation}
This circuit takes as input the control state $\ket{\vec{\kappa}}$, which is determined by the bit string representation of $m$.
By implementing the arithmetic formula for $\theta_{23}$, it computes the binary representation of the angle $\theta_{23}(\vec{\kappa})$ and stores it in the ancillary register.
The gate complexity of this circuit is $O(\poly(n,n_\theta))$.

Subsequently, the controlled-rotation gate
\begin{equation}
    U_{\mathrm{rot}}^{(2, 3)} = \sum_{\iota=0}^{2^{n_\theta}-1} R_x^{(2, 3)}(\theta_{23}^{(\iota)}) \otimes \ket{\iota}_a\bra{\iota}_a
\end{equation}
is applied, where $\theta_{23}^{(\iota)}$ denotes the angle corresponding to the integer $\iota$.
The action of this gate on the state prepared by the arithmetic circuit yields the desired transformation
\begin{equation}
    (I^{\otimes (n-2)} \otimes U_{\mathrm{rot}}^{(2, 3)}) \ket{\vec{\kappa}}\otimes\ket{\psi}_{\text{target}}\otimes \ket{\theta_{23}(\vec{\kappa})}_a
    = \ket{\vec{\kappa}}\otimes R_x^{(2, 3)}(\theta_{23}(\vec{\kappa}))\ket{\psi}_{\text{target}} \otimes \ket{\theta_{23}(\vec{\kappa})}_a,
\end{equation}
in which $\ket{\psi}_{\text{target}}$ is the two-qubit state encoding the spatial components.

Finally, to permit the reuse of the ancillary register, it is disentangled from the main system by applying the inverse of the computation circuit, $O_{23}^\dagger$, which performs the transformation $\ket{\vec{\kappa}} \otimes \ket{\theta_{23}(\vec{\kappa})} \mapsto \ket{\vec{\kappa}}  \otimes \ket{0}^{\otimes n_\theta}$.
This three-step sequence of computation, controlled rotation, and uncomputation yields the net evolution
\begin{equation}
    \ket{\vec{\kappa}} \otimes \ket{\psi}_{\text{target}} \otimes \ket{0}_a^{\otimes n_\theta}
    \mapsto \ket{\vec{\kappa}} \otimes R_x^{(2, 3)}(\theta_{23}(\vec{\kappa}))\ket{\psi}_{\text{target}} \otimes \ket{0}_a^{\otimes n_\theta},
\end{equation}
thereby realizing the operator defined in Eq.~\eqref{eq:opt_H23}.
The gate complexity of this entire process is $O(\poly(n, n_\theta))$.
An analogous procedure is employed for the remaining components, $\vec{L}_D$, $\vec{L}_{12}$, and $\vec{L}_{13}$.
We similarly apply a second-order TS decomposition to the operator $\ee^{\ii w_l\vec{H}_l / 2}$.

\begin{figure}[t!]
    \centering
    \includegraphics{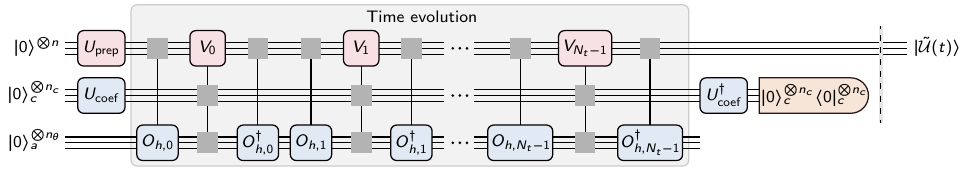}
    \caption{Quantum circuit for time evolution.
    Following the state preparation in Eq.~\eqref{eq:state_preparation_target}, the total evolution time $t$ is discretized into $N_t$ small time steps, implemented by the operator in Eq.~\eqref{eq:U_select}.
    Each constituent operator $V_{j,l}$ from Eq.~\eqref{eq:opt_each} is constructed using an oracle $O_{h,l}$ (e.g., $O_{23}$ in Eq.~\eqref{eq:O_23}) that computes the required rotation angles as arithmetic functions of the state index.
    The desired final state $\ket{\tilde{\mathcal{U}}(t)}$ is then obtained in the main register by applying the inverse operation $U_{\mathrm{coef}}^\dagger$ to the ancillary register and projecting it onto its ground state, with success probability $P \approx \norm{\tilde{\vec{\mathcal{U}}}(t)}_2^2$.
    The grey boxes denote control qubits or registers without specifying the control state ($\ket{0}$ or $\ket{1}$).
    }
    \label{fig:circuit_evolution}
\end{figure}

To isolate the desired solution, the ancillary register is disentangled from the main system via the inverse operation $U_{\mathrm{coef}}^\dagger$, which yields
\begin{equation}
    U_{\mathrm{coef}}^\dagger \otimes I^{\otimes n}: \frac{1}{\sqrt{\norm{\vec{c}}_1}} \sum_{j=0}^{M-1} \sqrt{c_j} \ket{j}_c \otimes U_{r_j}(t)\ket{\tilde{\mathcal{U}}(0)}
    \mapsto \frac{1}{\sqrt{\norm{\vec{c}}_1}} \sum_{j=0}^{M-1} \sqrt{c_j} (U_{\mathrm{coef}}^\dagger\ket{j}_c) \otimes U_{r_j}(t)\ket{\tilde{\mathcal{U}}(0)}.
\end{equation}
Projecting the resultant quantum state onto the ancillary basis state $\ket{0}_c^{\otimes n_c}$ gives
\begin{equation}\label{eq:meas_soln}
    \begin{aligned}
        \frac{1}{\sqrt{\norm{\vec{c}}_1}} \sum_{j=0}^{M-1} \sqrt{c_j} \bra{0}_c^{\otimes n_c}U_{\mathrm{coef}}^\dagger\ket{j}_c \otimes U_{r_j}(t)\ket{\tilde{\mathcal{U}}(0)}
        &= \bigg(\frac{1}{\sqrt{\norm{\vec{c}}}_1} \sum_{j'=0}^{M-1} \sqrt{c_{j'}}\bra{j'}_c\bigg) \bigg(\frac{1}{\sqrt{\norm{\vec{c}}_1}} \sum_{j=0}^{M-1} \sqrt{c_j} \ket{j}_c \otimes U_{r_j}(t)\ket{\tilde{\mathcal{U}}(0)} \bigg)
        \\
        &= \frac{1}{\norm{\vec{c}}_1} \sum_{j=0}^{M-1} c_j U_{r_j}(t)\ket{\tilde{\mathcal{U}}(0)},
    \end{aligned}
\end{equation}
which is precisely the desired solution $\tilde{\vec{\mathcal{U}}}(t)$, up to the normalization factor $1/\norm{\vec{c}}_1$.
Therefore, the final step is to measure the ancillary register.
If the measurement outcome is $\ket{0}_c^{\otimes n_c}$, the quantum state collapses to the state given by Eq.~\eqref{eq:meas_soln}.
The quantum circuit for state evolution is shown in Fig.~\ref{fig:circuit_evolution}.

\begin{figure}[t!]
    \centering
    \includegraphics{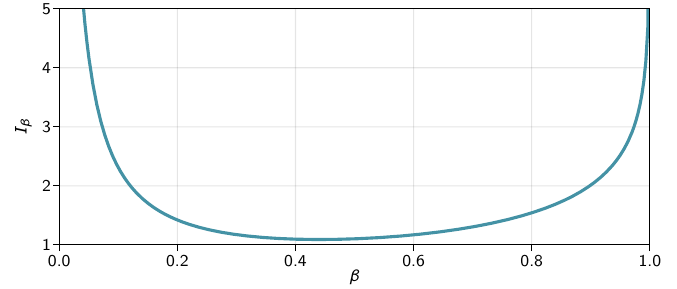}
    \caption{As a function of $\beta$, the integral $I_\beta$ in Eq.~\eqref{eq:P_success_est_1} attains a minimum value of $1.09$ at $\beta\approx 0.44$.}
    \label{fig:I_beta}
\end{figure}

The success probability of the measurement is given by the squared norm of the post-projection state as
\begin{equation}\label{eq:P_success_0}
    P = \bigg\|\frac{1}{\norm{\vec{c}}_1} \sum_{j=0}^{M-1} c_j U_{r_j}(t)\ket{\tilde{\mathcal{U}}(0)} \bigg\|_2^2
    = \frac{1}{\norm{\vec{c}}_1^2} \bigg\|\sum_{j=0}^{M-1} c_j U_{r_j}(t)\ket{\tilde{\mathcal{U}}(0)} \bigg\|_2^2.
\end{equation}
The normalization factor is approximated by the integral
\begin{equation}\label{eq:P_success_est_1}
    \norm{\vec{c}}_1 
    \approx \int_{-\infty}^\infty \bigg|\frac{\ee^{2^\beta - (1+\ii r)^{\beta}}}{2\pi (1-\ii r)} \bigg|\dif r 
    := I_\beta.
\end{equation}
A numerical evaluation of $I_\beta$, presented in Fig.~\ref{fig:I_beta}, reveals a minimum value of $1.09$ at $\beta\approx 0.44$, while remaining close to unity over a wide range of $\beta$.
% To maximize the post-selection success probability, we therefore choose $\beta=0.44$.
Then, substituting $I_\beta\approx 1$ and
\begin{equation}\label{eq:P_success_est_2}
    \bigg\|\sum_{j=0}^{M-1} c_j \vec{U}_{r_j}(t)\ket{\tilde{\vec{\mathcal{U}}}_0} \bigg\|_2^2
    \approx \norm{\tilde{\vec{\mathcal{U}}}(t)}_2^2
\end{equation}
into Eq.~\eqref{eq:P_success_0} yields the final success probability
\begin{equation}
    P \approx \norm{\tilde{\vec{\mathcal{U}}}(t)}_2^2.
\end{equation}
Consequently, for weak dissipation governed by $\tilde{\vec{L}}(t)$ or for short evolution times, the success probability $P \approx 1$, rendering the algorithm highly efficient.
Conversely, for significant dissipation or long evolution times, the success probability is severely diminished ($P \ll 1$).
Fortunately, the short-time dynamics of interest in RDT theory lie within this high-probability regime~\cite{Durbin2011_Statistical}, making the LCHS algorithm particularly suitable for this application.

\subsection{Statistics extraction through measurements}
The measurement complexity for observables requiring $2^n$ sampling points, e.g., the velocity-spectrum tensor and two-point autocovariance~\cite{Pope2000_Turbulent}, would nullify any quantum speedup, thereby precluding a quantum advantage even with efficient state preparation and evolution.
Consequently, it is crucial to identify global observables that characterize the flow's statistical properties using an $O(1)$ number of measurements.
In turbulence theory, prominent instances of such observables are the Reynolds stress tensor $R_{ij} = \widebar{u_iu_j}=\sum_{\vec{k}} \hat{u}_i(\vec{k})\hat{u}_j^*(\vec{k})$, and the energy spectrum $E(k)$, which for a discrete set of wavenumbers is defined as $E(k) = \sum_{|\vec{k}'|\approx k} |\hat{\vec{u}}(\vec{k}')|^2$.
Herein, we derive the measurement operators for the Reynolds stress and the energy spectrum, and provide efficient methods to obtain their expectation.

\subsubsection{Reynolds stress tensor}
With the state vector expanded as in Eq.~\eqref{eq:encode}, we construct a linear operator $\hat{R}_{ij}$ whose expectation value yields the Reynolds stress tensor, $\widebar{u_iu_j} = \ee^{2ct}\langle\tilde{\mathcal{U}} | \hat{R}_{ij} | \tilde{\mathcal{U}} \rangle$.
The corresponding Hermitian measurement operator is constructed as
\begin{equation}\label{eq:meas_Rij}
    \hat{R}_{ij}
    = \frac12 I^{\otimes n_N}\otimes (\ket{j}\bra{i} + \ket{i}\bra{j})
    = \frac12 \bigg(\sum_{m=0}^{N-1} \ket{m}\bra{m}\bigg) \otimes (\ket{j}\bra{i} + \ket{i}\bra{j})
    = \frac12 \sum_{m=0}^{N-1} (\ket{m, j}\bra{m, i} + \ket{m, i}\bra{m, j}).
\end{equation}
The evaluation of this expectation value, combining Eqs.~\eqref{eq:encode} and \eqref{eq:meas_Rij}, then gives
\begin{equation}
    \ee^{2ct}\langle\tilde{\mathcal{U}} | \hat{R}_{ij} | \tilde{\mathcal{U}} \rangle
    = \sum_{r=0}^{N-1} \sum_{s=0}^2 \hat{u}_s^*(\vec{\kappa}_r)\bra{r, s}
    \sum_{m=0}^{N-1} \ket{m, j}\bra{m, i}
    \sum_{p=0}^{N-1} \sum_{q=0}^2 \hat{u}_q(\vec{\kappa}_p)\ket{p, q}
    = \sum_{m=0}^{N-1} \hat{u}_i(\vec{\kappa}_m) \hat{u}_j^*(\vec{\kappa}_m)
    = \widebar{u_iu_j}.
\end{equation}
The operator $\hat{R}_{ij}$ is Hermitian by construction, ensuring its expectation value corresponds to a physical observable.
Its action is non-trivial only on the $n_d=2$ qubit component index register, rendering the unitary transformation required for measurement highly efficient.

We propose an efficient method to obtain the expectation $\langle\hat{R}_{ij}\rangle$, treating the cases $i\ne j$ and $i=j$ separately.
For the off-diagonal components where $i\ne j$, the possible measurement outcomes of the operator in Eq.~\eqref{eq:meas_Rij} are determined by its spectral decomposition.
This operator acts as an exchange operator on the subspace of the component index register spanned by $\{\ket{i}, \ket{j}\}$.
Within this subspace, we define the two orthonormal basis vectors
\begin{equation}\label{eq:def_psi+-}
    \ket{\psi_+} := \frac{1}{\sqrt{2}}(\ket{i} + \ket{j}), \quad
    \ket{\psi_-} := \frac{1}{\sqrt{2}}(\ket{i} - \ket{j}).
\end{equation}
For any state $\ket{m}$, the system's eigenvectors and eigenvalues are then classified into three categories.
First, eigenvectors of the form $\ket{m} \otimes \ket{\psi_+}$ possess the eigenvalue $\lambda_+ = 1/2$.
Second, eigenvectors of the form $\ket{m} \otimes \ket{\psi_-}$ correspond to the eigenvalue $\lambda_- = -1/2$.
Third, eigenvectors of the form $\ket{m} \otimes \ket{l_\perp}$, where $\ket{l_\perp}$ is any state orthogonal to both $\ket{i}$ and $\ket{j}$, have an eigenvalue of $\lambda_\perp = 0$.
Consequently, a single projective measurement of $\hat{R}_{ij}$ must yield one of these three eigenvalues.

\begin{figure}[t!]
    \centering
    \includegraphics{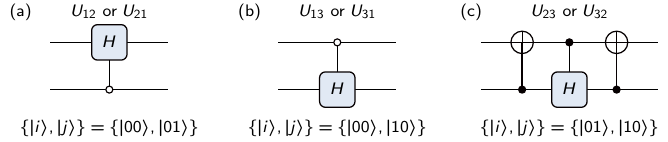}
    \caption{Quantum circuits for implementing the off-diagonal components of $U_{ij}$ in Eq.~\eqref{eq:U_ij}.}
    \label{fig:circuit_Uij}
\end{figure}

The measurement of $\hat{R}_{ij}$ is performed by a unitary transformation that maps its eigenbasis to the computational basis, followed by a projective measurement.
This unitary transformation, denoted $U_{ij}$, is constructed to act solely on the component index register.
The required mapping is defined by the conditions
\begin{equation}
    U_{ij} \ket{\psi_+} = \ket{i}, \quad
    U_{ij} \ket{\psi_-} = \ket{j}.
\end{equation}
Furthermore, $U_{ij}$ acts as the identity on any state $\ket{l_\perp}$ orthogonal to the subspace $\text{span}\{\ket{i}, \ket{j}\}$.
This operator is explicitly given by
\begin{equation}
    U_{ij} = \ket{i}\langle\psi_+| + \ket{j}\langle\psi_-| + \sum_{l_\perp \perp \{\ket{i}, \ket{j}\}} \ket{l_\perp}\bra{l_\perp}.
\end{equation}
Substituting the definitions for $\ket{\psi_+}$ and $\ket{\psi_-}$ in Eq.~\eqref{eq:def_psi+-}, we obtain
\begin{equation}\label{eq:U_ij}
    U_{ij} = \frac{1}{\sqrt{2}} (\ket{i}\langle i| + \ket{i}\langle j| + \ket{j}\langle i| - \ket{j}\langle j|) + \sum_{l_\perp \perp \{\ket{i}, \ket{j}\}} \ket{l_\perp}\bra{l_\perp}.
\end{equation}
Thus, the transformation is equivalent to a Hadamard gate on the subspace spanned by $\ket{i}$ and $\ket{j}$ and the identity on its orthogonal complement.
The explicit quantum circuits for implementing the off-diagonal components of $U_{ij}$ are provided in Fig.~\ref{fig:circuit_Uij}.

The measurement protocol for $i\ne j$ consists of applying the unitary operation $U_{ij}$ to the component index register of the prepared state $\ket{\tilde{\mathcal{U}}(t)}$, followed by a projective measurement in the computational basis.
A measurement outcome of $\ket{i}$ corresponds to a projection onto the eigenspace with eigenvalue $\lambda_+ = 1/2$, while an outcome of $\ket{j}$ corresponds to the eigenspace with eigenvalue $\lambda_- = -1/2$.
Any other orthogonal state outcome $\ket{l_\perp}$ signifies a projection onto the eigenspace associated with the eigenvalue $\lambda_\perp = 0$.
To estimate the expectation value $\langle\tilde{\mathcal{U}}(t)|\hat{R}_{ij}\ket{\tilde{\mathcal{U}}(t)}$, this procedure is repeated $N_M$ times, and the number of outcomes $N_i$, $N_j$, and $N_{l_\perp}$ are recorded.
The expectation value of the observable $\hat{R}_{ij}$ is then efficiently estimated as
\begin{equation}
    \langle \hat{R}_{ij} \rangle = \sum_{\lambda} \lambda \cdot p(\lambda) \approx \frac{1}{2}\times \frac{N_i}{N_M} -\frac{1}{2}\times \frac{N_j}{N_M} + 0\times \frac{N_k}{N_M} = \frac{N_i - N_j}{2N_M}.
\end{equation}

For the diagonal components where $i=j$, the operator in Eq.~\eqref{eq:meas_Rij} simplifies to the projector $\hat{R}_{ii}=\sum_{m=0}^{N-1} \ket{m,i}\bra{m,i}$.
Consequently, the measurement protocol reduces to a direct projection of the component index register onto the computational basis.
An outcome of $\ket{i}$ corresponds to a measurement value of 1, whereas any other outcome corresponds to a value of 0.
By repeating the measurement $N_M$ times and recording the number of counts $N_i$ for the state $\ket{i}$, the expectation value is efficiently estimated as $\langle \hat{R}_{ii} \rangle \approx N_i/N_M$.

\subsubsection{Energy spectrum}
Analogously to the Reynolds stress, the measurement operator for the energy spectrum is then constructed as
\begin{equation}
    \hat{E}(k) = \sum_{m\in S_k} \sum_{i=0}^2 \ket{m, i}\bra{m, i},
\end{equation}
where $S_k = \{m\in\{1,2,\cdots,N\} \mid |\vec{k}_m|= k\}$ denotes the set of indices for modes residing on the spherical shell of radius $k$.

To obtain statistically reliable results, the system is projected onto the computational basis $N_{M}$ times, yielding a set of outcomes $\{|m_j, i_j\rangle\}_{j=1}^{N_M}$.
These measurement outcomes are subsequently post-processed on a classical computer.
For each outcome, we determine if its mode index $m_j$ belongs to the target set $S_k$ by evaluating the condition $|\vec{k}_{m_j}| = k$.
The total number of outcomes $N_k$ satisfying this condition across the $N_M$ trials is then counted as
\begin{equation}
    N_k = \sum_{j=1}^{N_M} \mathbb{I}(m_j \in S_k),
\end{equation}
where $\mathbb{I}(\cdot)$ denotes the indicator function.
The expectation value of the observable $\hat{E}(k)$ is then estimated by the frequency of successful outcomes, $\langle \hat{E}(k) \rangle \approx N_k/N_M$.
The statistical uncertainty of this estimator is given by $\sigma = \sqrt{\langle\hat{E}(k)\rangle(1 - \langle\hat{E}(k)\rangle) / N_M}$, which decreases with an increasing number of measurements according to the central limit theorem.

\subsection{Complexity analysis}
We summarize the time complexity of each step and compare the total complexity to its classical counterpart.
The simulation requires $n=\lceil \log_2 N \rceil+2$ qubits to encode the velocity field, $n_c=\lceil\log_2 M\rceil$ qubits for the LCU coefficients with $M = O(t \log^{1+1/\beta}(1/\varepsilon))$, and $n_\theta$ ancillary qubits for arithmetic operations.
The gate complexity for preparing the initial state by encoding the initial velocity and the LCU coefficients to a precision $\varepsilon$ is $\mathcal{C}_{\mathrm{prep}}=O(\poly(n,n_\theta), n_c \log(1/\varepsilon))$.
Given a target precision $\varepsilon$ of $\ket{\tilde{\mathcal{U}}}$, the time evolution stage exhibits a gate complexity of $\mathcal{C}_{\mathrm{evo}}=O(\poly(n,n_c,n_\theta)t\log(1/\varepsilon))$.

% Let $\varepsilon_m$ denote the target additive precision for the estimation of turbulent statistics.
In the measurement stage, which involves a pre-measurement circuit of negligible depth, to achieve the target precision $\varepsilon_m$ for turbulent statistics, the procedure must be repeated $N_M$ times. 
With standard projective measurements, the required number of samples is $N_M = O(1/\varepsilon_m^2)$, according to the central limit theorem.
Note that turbulent statistics are quadratic forms of all state-vector components, the error $\varepsilon_m$ is thus independent of $N$ and satisfies $\varepsilon_m \gg \varepsilon$.
Employing quantum amplitude estimation (QAE)~\cite{Brassard2002_Quantum} can quadratically reduce the measurement cost. 
The query complexity for QAE to achieve precision $\varepsilon_m$ is $O(1/\varepsilon_m)$, offering an advantage over the classical Monte Carlo sampling.

Consequently, the total time complexity of the end-to-end quantum algorithm is
\begin{equation}
    \mathcal{C}_{\mathrm{qc}} 
    = N_M(\mathcal{C}_{\mathrm{prep}} + \mathcal{C}_{\mathrm{evo}})
    = O\bigg(\poly(n,n_c,n_\theta) \frac{t\log(1/\varepsilon)}{\varepsilon_m}\bigg).
\end{equation}
% The algorithm thus runs in polynomial time for $N \gtrsim O(1/\varepsilon^{11/4})$.
For comparison, the computational complexity of classical simulation is $\mathcal{C}_{\mathrm{cc}} = O(Nt\|\tensor{A}\|_2)$.
Since $\norm{\tensor{A}}_2$ is an $O(1)$ constant in rapidly distorted turbulence, the resulting quantum speedup is then
\begin{equation}\label{eq:speedup}
    \mathcal{S} = \frac{\mathcal{C}_{\mathrm{cc}}}{\mathcal{C}_{\mathrm{qc}}}
    = O\bigg(\frac{N \varepsilon_m}{\poly(\log N) \log(1/\varepsilon)} \bigg).
\end{equation}
This result indicates a practical quantum speedup on a sufficiently large computational grid, provided that $N \gtrsim O(1/\varepsilon_m)$.

The use of high-order numerical schemes is critical to the algorithm's efficiency.
This choice directly yields the polylogarithmic scaling of the LCU coefficient count, $M = O(t \log^{1+1/\beta}(1/\varepsilon))$.
In contrast, a low-order discretization, such as the trapezoidal rule, would yield a polynomial scaling of $M$ with $1/\varepsilon$, rendering the algorithm computationally prohibitive.
Thus, for LCU-based quantum simulations, the implementation overhead of high-order methods is substantially outweighed by the exponential savings in required quantum operations.

\section{Results}\label{sec:results}
\subsection{Numerical setup}
We numerically validate the proposed algorithm for three-dimensional rapidly distorted turbulence with finite mean vorticity.
This flow is characterized by the linear velocity profile $U_1=\mathcal{S}x_2$ ($\mathcal{S} > 0$) and a constant mean vorticity $\varOmega_3=-\mathcal{S}$, corresponding to the mean velocity-gradient tensor
\begin{equation}
    \tensor{A} = \begin{bmatrix}
        0 & \mathcal{S} & 0 \\
        0 & 0 & 0 \\
        0 & 0 & 0
    \end{bmatrix}.
\end{equation}
% The maximum singular value of $\tensor{A}$ is thus $\sigma_{\max}(\tensor{A})=|\mathcal{S}|$.
The tensor $\tensor{L}_d(\vec{\kappa})$ in Eq.~\eqref{eq:L_d} then simplifies to
\begin{equation}
    \tensor{L}_d = \mathcal{S}\begin{bmatrix}
        0 & \frac{\kappa_1^2}{|\vec{\kappa}|^2} - \frac12 & 0 \\
        \frac{\kappa_1^2}{|\vec{\kappa}|^2} - \frac12 & \frac{2\kappa_1\kappa_2}{|\vec{\kappa}|^2} & \frac{\kappa_1\kappa_3}{|\vec{\kappa}|^2} \\
        0 & \frac{\kappa_1\kappa_3}{|\vec{\kappa}|^2} & 0
    \end{bmatrix}.
\end{equation}
The maximum eigenvalue of $\tensor{L}_d$ is
\begin{equation}
    \lambda_{\max}(\tensor{L}_d) = \frac{\mathcal{S}}{2}\bigg(\frac{2\kappa_1\kappa_2}{|\vec{\kappa}|^2} + \sqrt{\bigg(\frac{2\kappa_1 \kappa_2}{|\vec{\kappa}|^2} \bigg)^2 + 4\bigg[\bigg(\frac{\kappa_1^2}{|\vec{\kappa}|^2} - \frac12 \bigg)^2 + \bigg(\frac{\kappa_1\kappa_3}{|\vec{\kappa}|^2} \bigg)^2 \bigg]} \bigg)
    \le \mathcal{S},
\end{equation}
reaching its maximum value $\mathcal{S}$ when $\kappa_1=\kappa_2$ and $\kappa_3=0$.
Consequently, we set the parameter $c=\mathcal{S}$.
The initial velocity is generated from Eq.~\eqref{eq:IC_vel} with the model spectrum $E_0(|\vec{k}|)=k^{-5/3}f_L(kL)f_\eta(k\eta)$, which ensures that energy is distributed over a wide range of scales.
The parameters are set to $L=2\pi$, $\eta=0.1$, $c_L=6.78$, $c_\eta=0.4$, and $\beta=5.2$, in accordance with commonly used values~\cite{Pope2000_Turbulent}.

\begin{figure}[t!]
    \centering
    \includegraphics{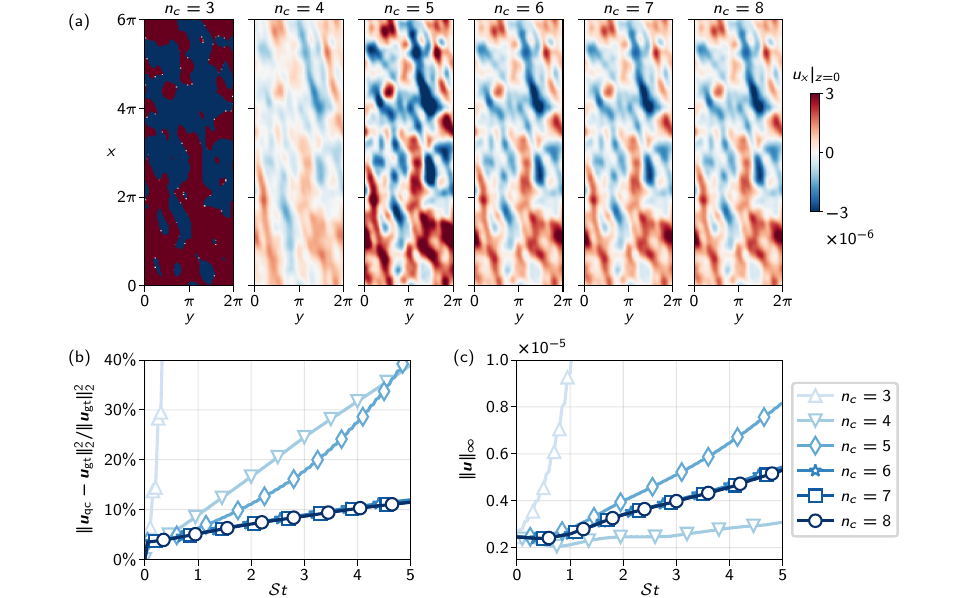}
    \caption{Effect of the ancilla-qubit number $n_c$ on the LCHS simulation.
    (a) Fluctuating streamwise velocity in the $z=0$ plane at $\mathcal{S}t=5$, for $n_c=3,4,\cdots,8$.
    The corresponding relative $L^2$ velocity error and maximum velocity are plotted in (b) and (c), respectively.
    These quantitative results demonstrate convergence for $n_c=6$.}
    \label{fig:compare_nc}
\end{figure}

The flow is considered in a periodic domain $[0,6\pi]\times [0,2\pi]\times [0,2\pi]$, which is elongated in the mean-flow direction.
We set the mean shear $\mathcal{S} = 10$.
The domain is discretized on a uniform $256\times 64\times 64$ grid, encoded by $n_x=8$, $n_y=6$, and $n_z=6$ qubits for the respective spatial directions, totaling $n=n_x+n_y+n_z+2=22$ qubits.
The trapezoidal rule is employed for the numerical implementation instead of Gaussian quadrature, motivated by the superior performance of low-order schemes on small-scale problems, which stems from their favorable node symmetry.
A detailed comparison between these two integration schemes is presented in \ref{app:compare_low_high}.
The simulation is evolved for $N_t=100$ time steps to a final time $t=0.5$, with additional $n_c$ qubits for encoding the LCU coefficients and a truncated boundary of $R=2^{n_c-1}$.
A convergence study for $n_c=3, \cdots, 8$ indicates that the LCHS solution converges for $n_c=6$, as shown in Fig.~\ref{fig:compare_nc}.
Besides, we set the parameter $\beta=0.8$ as a compromise between the post-selection success probability and numerical accuracy.

For the benchmarking instance, we implement the LCHS approach on a classical computer as an analog of the proposed quantum algorithm.
Our implementation focuses on the time-evolution stage, efficiently realizing the linear combination of unitaries in Eq. \eqref{eq:U_select} via vectorized operations.
In contrast, the initial state preparation and measurement are performed by direct classical read-in and read-out.

The quantum circuit depth is estimated from the number of required TS decompositions.
To balance accuracy and computational cost, a second-order TS scheme is employed for two nested levels of decomposition.
For each of the $N_t = 100$ time steps, an outer decomposition separates the non-commuting operators $\ee^{\ii w_l\tensor{H}_l}$ and $\ee^{\ii r_jw_l\tensor{L}_l}$, yielding a three-term sequence per Eq.~\eqref{eq:opt_each}.
Subsequently, an inner decomposition expands each exponential, such as $\ee^{\ii r_jw_l\tensor{L}_l}$, into seven elementary operators corresponding to its four non-commuting components $(\tensor{L}_D, \tensor{L}_{12}, \tensor{L}_{13}, \tensor{L}_{23})$ in Eq.~\eqref{eq:evolution_element}.
The total circuit depth is therefore approximately $100\times 3 \times 7 = 2100$.
This constitutes a minimal circuit scale for simulating a non-trivial turbulence problem.
The present work consequently determines the minimum quantum resources required to simulate a physically meaningful, albeit simplified, turbulent phenomenon.
Such a result establishes a quantitative baseline for benchmarking future, more complex quantum fluid simulations.

\begin{figure}[t!]
    \centering
    \includegraphics{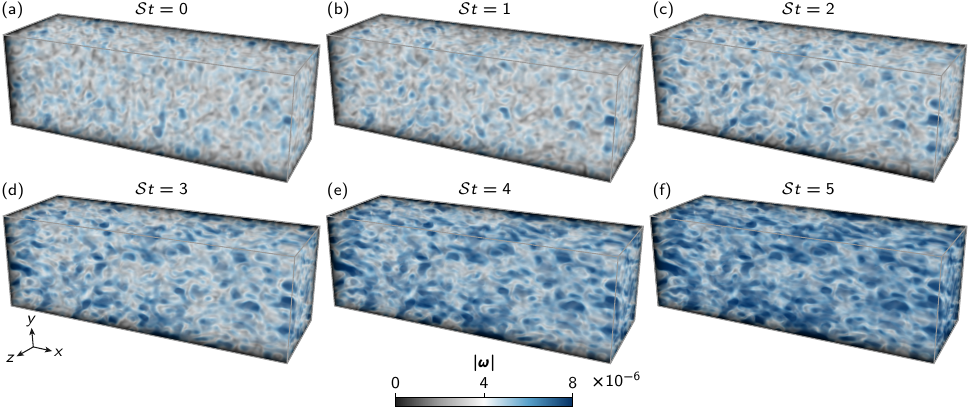}
    \caption{Volume rendering of the fluctuating vorticity magnitude $|\vec{\omega}|$ in rapidly distorted turbulence obtained by quantum simulation with $n_c=6$ ancilla qubits.
    Energy extracted from the mean flow intensifies the fluctuating vorticity through the streamwise stretching of vortical structures.}
    \label{fig:vor_render_evolution}
\end{figure}

\subsection{LCHS simulation of rapidly distorted homogeneous shear turbulence}
Here, we present the LCHS simulation results, benchmarking them against the ground-truth solution.

The volume renderings of vorticity magnitude obtained by quantum simulation illustrate the organization of turbulent structures by a mean shear, as shown in Fig.~\ref{fig:vor_render_evolution}.
Initially, the flow consists of disorganized and blob-like vortical structures, approximating the state of nearly homogeneous isotropic turbulence.
The mean velocity gradient subsequently stretches, tilts, and reorients these structures along the principal strain-rate axis.
This process elongates the vortex tubes and contracts their cross-sections, thereby intensifying the core vorticity as dictated by the Helmholtz vorticity theorem.
Visually, this corresponds to the emergence of elongated, sheet-like coherent structures with amplified vorticity magnitude in Fig.~\ref{fig:vor_render_evolution}(f).
This evolution demonstrates the extraction of kinetic energy from the mean flow, which transforms the disordered initial state into a highly organized and anisotropic one.

\begin{figure}[t!]
    \centering
    \includegraphics{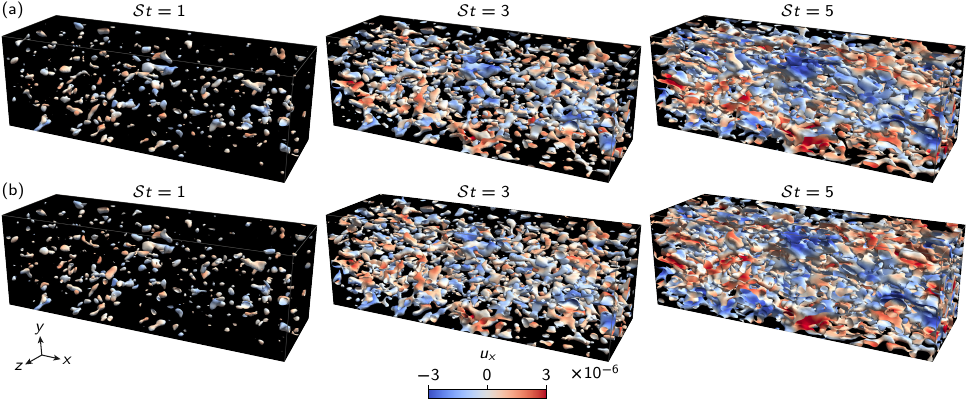}
    \caption{Evolution of the fluctuating vorticity isosurface $|\vec{\omega}|=8\times 10^{-6}$, colored by the streamwise velocity fluctuation $u_x$ at $\mathcal{S}t=1, 3,$ and 5.
    The quantum simulation results in (a), obtained with $n_c=6$ ancilla qubits, are nearly indistinguishable from the ground-truth solution in (b).}
    \label{fig:iso-vor}
\end{figure}

Figure~\ref{fig:iso-vor} compares the instantaneous vorticity isosurfaces from the quantum simulation with the ground-truth solution at $\mathcal{S}t=1, 3,$ and 5.
The excellent vortex-structure agreement confirms the capability of the quantum algorithm to capture the essential flow dynamics.
The visualization reveals a progressive organization of initially isotropic bubble-like vortical structures into anisotropic bands inclined relative to the streamwise direction.
This alignment is governed by the principal axis of positive strain of the mean flow, which tilts, reorients, and stretches the nascent vortical structures.
% The superimposed streamlines kinematically confirm this process, forming highly elongated eddies with strong vorticity.

\begin{figure}[t!]
    \centering
    \includegraphics{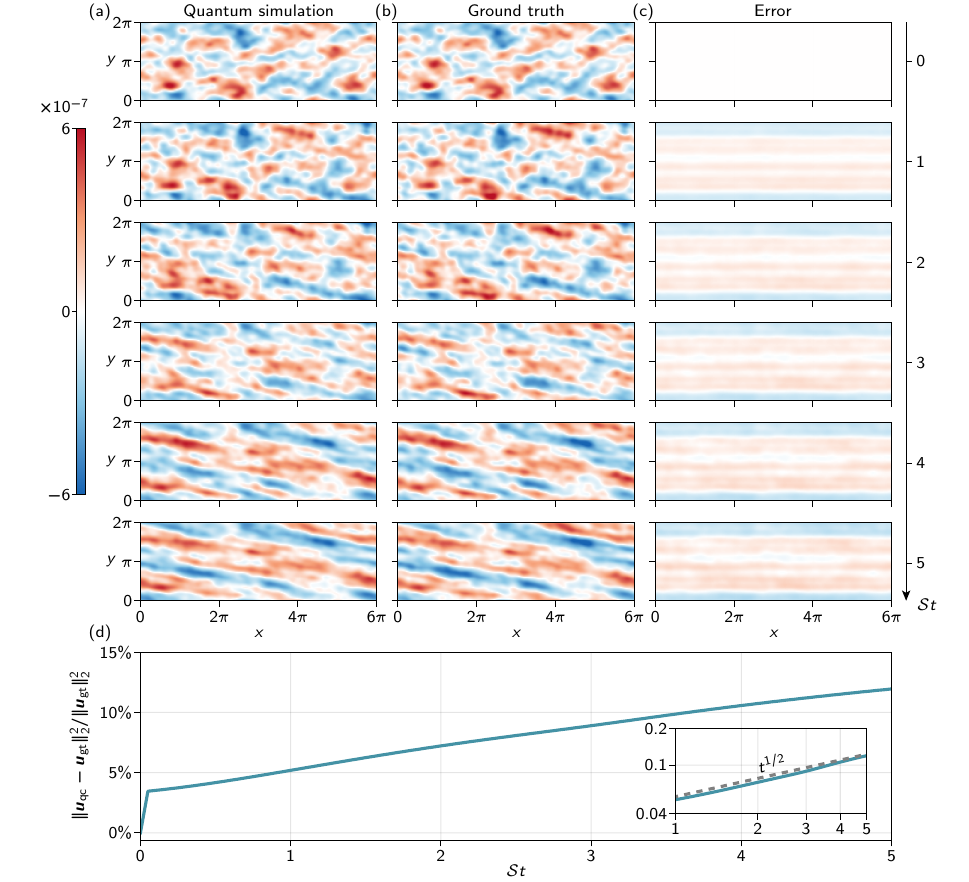}
    \caption{Evolution of the spanwise-averaged streamwise velocity $\langle u_x\rangle_z$ at $\mathcal{S}t=0,1,\cdots,5$.
    The quantum simulation results $\vec{u}_{\mathrm{qc}}$ in (a), obtained with $n_c=6$ ancilla qubits, demonstrate excellent agreement with the ground-truth solution $\vec{u}_{\mathrm{gt}}$ in (b), with the difference displayed in (c).
    (d) The relative $L^2$ error $\norm{\vec{u}_{\mathrm{qc}} - \vec{u}_{\mathrm{gt}}}_2^2 / \norm{\vec{u}_{\mathrm{gt}}}_2^2$ grows stably at a rate of $\sim t^{1/2}$ at later times.}
    \label{fig:mean_velx}
\end{figure}

The spanwise-averaged streamwise fluctuating velocity $\langle u_x\rangle_z$ from the quantum simulation in Fig.~\ref{fig:mean_velx}(a) is nearly identical to the ground-truth solution in Fig.~\ref{fig:mean_velx}(b).
The relative $L^2$ error, shown in Fig.~\ref{fig:mean_velx}(d), grows stably at a rate of $\sim t^{1/2}$ at later times.
This error growth stems from the truncation error of the second-order TS decomposition, which is chosen as a compromise between computational cost and accuracy.

\begin{figure}[t!]
    \centering
    \includegraphics{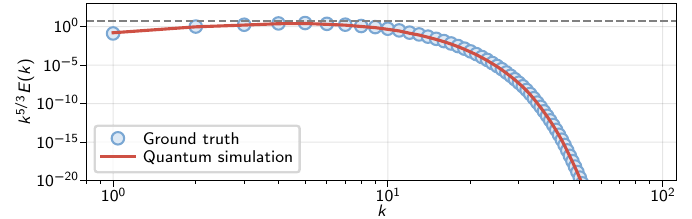}
    \caption{Compensated spectrum $k^{5/3}E(k)$ of the velocity fluctuations at $\mathcal{S}t=5$.
    The quantum simulation result, obtained with $n_c=6$ ancilla qubits, is indistinguishable from the ground truth solution, revealing a narrow inertial range.}
    \label{fig:Ek}
\end{figure}

The spanwise-averaged streamwise fluctuating velocity exhibits alternating high- and low-speed streaks at $\mathcal{S}t=5$, which are elongated in the streamwise direction and tilted within the $x$-$y$ plane, with the energy spectrum shown in Fig.~\ref{fig:Ek}.
These structures are a manifestation of the linear lift-up mechanism in RDT.
In this process, the mean shear $\mathcal{S}$ tilts the initial fluctuating vorticity to generate a streamwise component $\omega_x$.
The resulting streamwise vortices induce a normal velocity $u_y$, which in turn advects the mean momentum.
Specifically, upwash motion ($u_y>0$) lifts low-momentum fluid, creating a low-speed streak ($u_x<0$), whereas downwash motion ($u_y<0$) advects high-momentum fluid, forming a high-speed streak ($u_x>0$).

The tilted geometry of the streaks is dictated by the orientation of the underlying vortical structures.
This orientation results from a competition between the mean strain, which stretches vortices along its $45^\circ$ principal axis, and the mean rotation, which shears them toward the streamwise direction.
This interplay yields a net vortex tilt of less than $45^\circ$, as shown in Fig.~\ref{fig:mean_velx}(a) at $\mathcal{S}t=5$.
The high- and low-speed streaks generated by these tilted vortices therefore inherit the same inclination.
These results thus demonstrate that linear dynamics alone are sufficient to organize an initially disordered field into the coherent streaks characteristic of shear turbulence.

\begin{figure}[t!]
    \centering
    \includegraphics{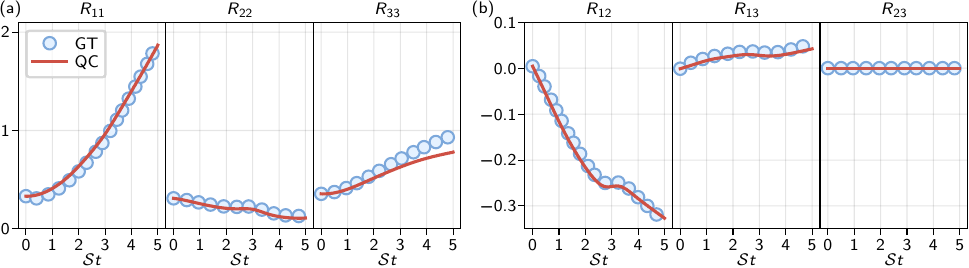}
    \caption{Evolution of the Reynolds stresses: (a) normal stresses and (b) shear stresses.
    The quantum simulation results (red lines), obtained with $n_c=6$ ancilla qubits, are in agreement with the ground-truth solution (blue circles), with the notable exception of the $R_{33}$ component.
    This underestimation is attributed to the coarse grid's insufficient resolution of the pressure-strain energy redistribution.}
    \label{fig:Reynolds_stress}
\end{figure}

In Fig.~\ref{fig:Reynolds_stress}, the evolution of the Reynolds stress tensor provides a statistical description of these physical processes, quantifying the systematic extraction of energy from the mean flow by the fluctuations.
The development of the Reynolds shear stress $R_{12}$ drives this energy extraction.
Specifically, the lift-up mechanism establishes a strong negative correlation between the streamwise $u_x$ and normal $u_y$ velocity fluctuations, as upwash ($u_y>0$) advects low-speed fluid ($u_x<0$) and downwash ($u_y<0$) advects high-speed fluid ($u_x>0$).
Consequently, the product $u_x u_y$ is predominantly negative, yielding a significant negative Reynolds shear stress $R_{12}$.
This negative shear stress in turn drives the growth of the streamwise normal stress $R_{11}$ through its production term, $P_{11}=-2\mathcal{S}R_{12}$.
Since $R_{12}<0$ and $\mathcal{S}>0$, this term is large and positive, signifying a continuous transfer of energy from the mean flow to the streamwise fluctuations.
In contrast, the growth mechanisms for the other normal stresses, $R_{22}$ and $R_{33}$, are different.
Their direct production terms are identically zero, so their growth depends entirely on energy redistribution by the pressure-strain correlation term.
This term channels energy from the dominant $R_{11}$ component to $R_{22}$ and $R_{33}$, which drives the turbulence towards a more isotropic state.
The variation of $R_{22}$ and $R_{33}$ is thus an indirect, lagging process that requires high numerical resolution to capture accurately.
The quantum simulation with an insufficient number of time steps $N_t$, therefore, visibly underestimates $R_{33}$ in Fig~\ref{fig:Reynolds_stress}(a).
This entire sequence of energy production and redistribution culminates in the characteristic anisotropy of shear turbulence, where $R_{11}$ is the largest component.

\section{Conclusions}\label{sec:conclusions}
We propose a quantum algorithm to simulate the dynamics of rapidly distorted turbulence, based on the RDT framework and the LCHS method.
The algorithm integrates the efficient quantum preparation of an initial turbulent state, the time evolution of fluctuating fields governed by the RDT equations, and the direct quantum measurement of statistical quantities such as the Reynolds stresses.
This integrated measurement protocol circumvents the need for a full state tomography, offering a more direct path to extracting physical observables.
For a sufficiently large number of grid points, our analysis indicates a practical quantum speedup over the classical turbulence simulation methods, as quantified by Eq.~\eqref{eq:speedup}.

We quantify the minimal cost of simulating a non-trivial turbulence on a quantum computer. A short-time evolution of three-dimensional rapidly distorted turbulence on a $256\times 64\times 64$ grid requires a quantum simulation using a 26- to 28-qubit quantum computer and 2100 Trotter steps. 
%The classical simulation results are in excellent agreement with the ground-truth solution.
The numerical results faithfully reproduce key physical phenomena, including the generation of high- and low-speed streaks by the lift-up mechanism, the development of Reynolds stress anisotropy, and the characteristic tilting of coherent structures.
%This high-fidelity simulation of rapidly distorted turbulence demonstrates the potential of our quantum algorithm for physically meaningful applications.
This result establishes a foundational benchmark for the quantum computational cost of fluid dynamics, advancing the field from theoretical algorithms to concrete resource estimations for practical problems.

However, the present approach has several limitations.
First, from an implementation perspective, the LCHS method's reliance on numerous multi-controlled gates results in deep quantum circuits.
Its execution on near-term noisy intermediate-scale quantum devices is therefore challenging~\cite{Meng2025_Challenges}, despite its theoretical quantum speedup over classical algorithms for large-scale problems.
Considering the substantial computational cost of simulating the quantum algorithm classically, we do not test its robustness to noise.
Second, from a physical modeling standpoint, our study is constrained by the RDT framework.
This linear theory captures only the initial stage of turbulent evolution, neglecting nonlinear processes such as the energy cascade and vortex-vortex interactions.
Despite this limitation, RDT isolates key mechanisms such as the lift-up effect, thereby providing a vital conceptual bridge.
Insights derived from its quantum representation may guide the encoding of complex nonlinear interactions in the full NS equations.
Third, the practical preparation of high-fidelity initial states for realistic turbulent flows remains a significant challenge~\cite{Meng2025_Geometric}.
A further key challenge is the extension of the state-preparation algorithm from idealized rapidly distorted homogeneous shear flow to more complex, inhomogeneous flows.

This work provides a foundation for the quantum simulation of more complex turbulent phenomena.
A promising future direction involves generalizing the RDT framework to a system of stochastic differential equations.
Within this stochastic framework, the statistical effects of nonlinear terms, such as vortex interactions, can be parameterized by a calibrated random forcing term~\cite{Farrell1993_Stochastic}.
This approach could bridge the gap between the linear RDT-LCHS framework and fully developed turbulence, enabling the simulation of statistically stationary states.
Such a development would allow quantum algorithms to address more realistic turbulence scenarios.
% A primary long-term objective is to extend this framework to the full nonlinear fluid dynamics.
% Near-term research could focus on more hardware-efficient methods, such as variational or hybrid quantum-classical approaches, to mitigate the challenge of circuit depth on NISQ-era devices.
% Ultimately, the maturation of these quantum methods promises a new computational paradigm for addressing classically intractable problems in fields ranging from aerospace engineering to climate science.

\section*{Declaration of competing interest}
The authors declare no conflict of interest. 

\section*{Acknowledgments}
ZM and GH acknowledge support from NSFC the Excellence Research Group Program for multiscale problems in nonlinear mechanics (Grant No.~12588201). JPL acknowledges support from Innovation Program for Quantum Science and Technology (Grant No.~2024ZD0300502), start-up funding from Tsinghua University and Beijing Institute of Mathematical Sciences and Applications.

\appendix

\section{Trapezoidal rule versus the Gaussian quadrature for the LCHS implementation}\label{app:compare_low_high}
\renewcommand{\thefigure}{\Alph{section}\arabic{figure}}
\setcounter{figure}{0}

\begin{figure}[t!]
    \centering
    \includegraphics{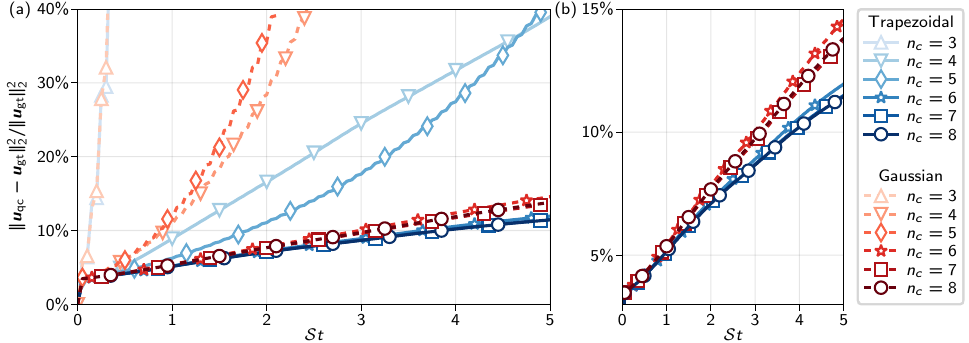}
    \caption{Relative $L^2$ error of the fluctuating velocity with $n_c=3,4,\cdots,8$ ancilla qubits, comparing the trapezoidal rule with Gaussian quadrature in the LCHS implementation.
    For the latter, we fix $R=32$ and $Q=4$, while varying $I_R=2,4,\cdots, 64$.
    (b) An enlarged view of the converged results for $n_c\ge 6$.}
    \label{fig:compare_velL2error_scheme}
\end{figure}

\begin{figure}[t!]
    \centering
    \includegraphics{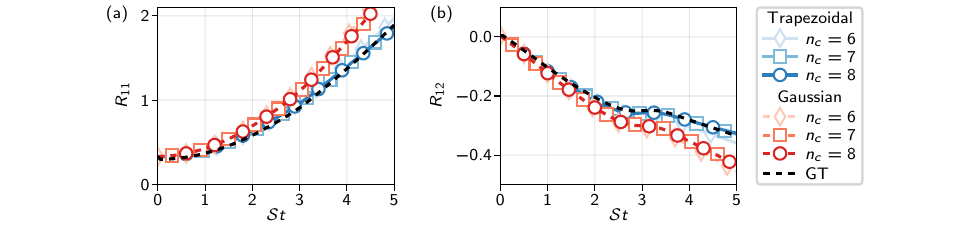}
    \caption{Reynolds stresses (a) $R_{11}$ and (b) $R_{12}$ with $n_c=6,7,8$ ancilla qubits, comparing the trapezoidal rule with Gaussian quadrature in the LCHS implementation.}
    \label{fig:compare_Rij_scheme}
\end{figure}

We compare the performance of the LCHS method using two distinct integral discretization schemes, high-order Gaussian quadrature and the low-order trapezoidal rule, for $n_c=3,4,\cdots,8$ ancilla qubits.
For the Gaussian quadrature, we fix $R=32$ and $Q=4$, while varying the number of subintervals $I_R=2,4,\cdots,64$.
For the trapezoidal rule, the number of intervals is set to $R=2^{n_c-1}$.

Figure~\ref{fig:compare_velL2error_scheme} compares the relative $L^2$ error of the fluctuating velocity for the two integration schemes.
Although the results from both methods converge for $n_c \ge 6$, the trapezoidal rule exhibits superior performance, yielding a smaller error for a given number of ancilla qubits.
Moreover, the Reynolds stresses $R_{11}$ and $R_{12}$ obtained using the trapezoidal rule are indistinguishable from the ground-truth solution, whereas the results from the Gaussian quadrature display a clear deviation in Fig.~\ref{fig:compare_Rij_scheme}.

The superior performance of the low-order trapezoidal rule is attributed to its inherent nodal symmetry, as its equally spaced nodes provide a symmetric sampling of the integrand.
In contrast, high-order Gaussian quadrature utilizes asymmetrically placed nodes optimized for polynomial approximation.
Within the LCHS framework, this asymmetry can result in a less balanced representation of the simulated operator, an effect that is particularly pronounced in small-scale systems sensitive to the choice of discretization.

\bibliographystyle{elsarticle-num.bst}
\bibliography{main.bib}

\end{document}